\documentclass{article}

\usepackage{microtype}
\usepackage{graphicx}
\usepackage{booktabs} % for professional tables
\usepackage{caption}
\usepackage{subcaption} % Required for creating subfigures
\usepackage{tabularx}  % For large tables
\usepackage{hyperref}
\usepackage{tikz}

\definecolor{sbsblue}{HTML}{1f77b4}
\definecolor{sbsorange}{HTML}{ff7f0e}
\definecolor{sbsgreen}{HTML}{2ca02c}
\definecolor{sbsred}{HTML}{d62728}

\usetikzlibrary{cd,arrows,calc,shapes,positioning}
\tikzstyle{obs} = [circle,fill=white,draw=black,inner sep=0pt,minimum size=18pt,font=\fontsize{10}{10}\selectfont,node distance=1,thick]
\tikzstyle{latent} = [obs,dotted]

\usepackage[accepted]{icml2024}

\usepackage{amsmath}
\usepackage{amssymb}
\usepackage{mathtools}
\usepackage{amsthm}

\usepackage[capitalize,noabbrev]{cleveref}

\theoremstyle{plain}

\theoremstyle{definition}

\theoremstyle{remark}

\usepackage[textsize=tiny]{todonotes}

\icmltitlerunning{Position: AI/ML Influencers Have a Place in the Academic Process}

\begin{document}

\twocolumn[
\icmltitle{Position: AI/ML Influencers Have a Place \\ in the Academic Process}



\icmlsetsymbol{equal}{*}

\begin{icmlauthorlist}
\icmlauthor{Iain Xie Weissburg}{equal,ece}
\icmlauthor{Mehir Arora}{equal,cs}
\icmlauthor{Xinyi Wang}{cs}
\icmlauthor{Liangming Pan}{cs}
\icmlauthor{William Yang Wang}{cs}
\end{icmlauthorlist}

\icmlaffiliation{cs}{Department of Computer Science, University of California, Santa Barbara, USA}
\icmlaffiliation{ece}{Department of Electrical and Computer Engineering, University of California, Santa Barbara, USA}

\icmlcorrespondingauthor{Iain Xie Weissburg}{ixw@ucsb.edu}

\icmlkeywords{Machine Learning, ICML}

\vskip 0.3in ]



\printAffiliationsAndNotice{\icmlEqualContribution} % otherwise use the standard text.

\begin{abstract}

As the number of accepted papers at AI and ML conferences reaches into the thousands, it has become unclear how researchers access and read research publications. In this paper, we investigate the role of social media influencers in enhancing the visibility of machine learning research, particularly the citation counts of papers they share. We have compiled a comprehensive dataset of over 8,000 papers, spanning tweets from December 2018 to October 2023, alongside controls precisely matched by 9 key covariates. Our statistical and causal inference analysis reveals a significant increase in citations for papers endorsed by these influencers, with median citation counts 2-3 times higher than those of the control group. Additionally, the study delves into the geographic, gender, and institutional diversity of highlighted authors. Given these findings, we advocate for a responsible approach to curation, encouraging influencers to uphold the journalistic standard that includes showcasing diverse research topics, authors, and institutions.

\end{abstract}

\section{Introduction}
\label{sec:intro}

\begin{figure}
    \centering
    \includegraphics[width=\columnwidth]{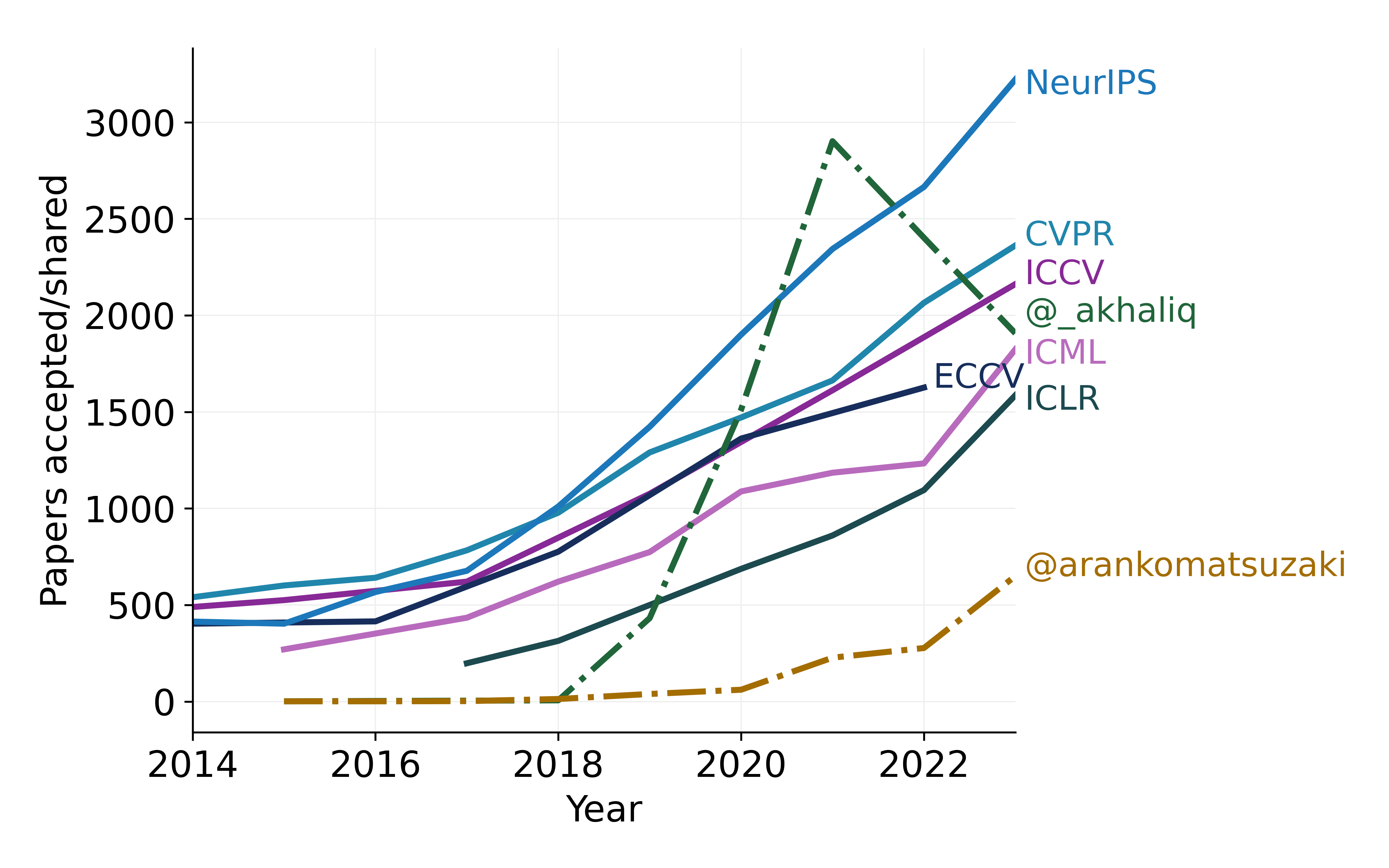}
    \caption{The number of papers accepted to top AI/ML conferences (solid) and shared by influencers (dashed) from 2014-2023 \cite{Li2023ConferenceAR}.}
    \label{fig:paper_acceptance}
\end{figure}

In the evolving landscape of artificial intelligence and machine learning (AI/ML), the exponential increase in conference papers, as depicted in Figure~\ref{fig:paper_acceptance}, alongside rapid technological advancements, has significantly transformed the dissemination of scholarly knowledge. A notable aspect of this transformation is the practice of online preprint sharing, with platforms like ArXiv becoming particularly prominent within the AI/ML community. This phenomenon allows for early access to research, often months before official publication, raising questions about these papers' evolving relevance and reception in traditional academic forums. Most notably, how do people select papers to read in the online age?

This paper aims to explore the changing dynamics of academic discourse within the AI/ML community, emphasizing the constructive role of social media in addressing the challenges posed by the sheer volume of literature. We study the class of $\mathbb{X}$ (formerly Twitter) influencers. Specifically, we require that influencers (1) share freshly posted preprints from ArXiv, (2) regularly post about papers, (3) do not focus on self-promotion, and (4) have a large audience. AK (@\_akhaliq) and Aran Komatsuzaki (@arankomatsuzaki) are the only influencers found to satisfy our requirements. 

Acknowledging the vital function of these influencers as curators, we examine the impact of their endorsements on the citation counts of shared papers. However, we also underscore the importance of maintaining a balanced research ecosystem. An over-reliance on a select group of curators may inadvertently skew the research landscape, emphasizing certain topics or perspectives over others. Therefore, we advocate for a responsible approach to curation, encouraging influencers to maintain journalistic standards, showcasing diverse research topics, authors, and institutions.

\paragraph{Contributions.} To this end, we provide the following three main contributions to the discussion about the increasing impact of academic social media figures in the AI/ML domain:

$\bullet$ \textbf{Comprehensive Dataset with Control Samples. } We collect a dataset of shared papers from AK and Komatsuzaki containing key paper information. We select control samples through precise matching, using several paper and author covariates. 

$\bullet$ \textbf{Thorough Analysis of Citations and Demographics. } Our comprehensive analysis aims to determine if papers these influencers share receive higher citation counts than non-shared papers. We conduct statistical and causal inference analysis and find significant evidence for a direct effect. We then explore the geographic and gender distributions of influencer-shared papers and compare them to reference points from the community.

$\bullet$ \textbf{Proposals for Future AI/ML Information Sharing. } Finally, we propose that the academic community engage in a future discussion on evolving the conference system. We want to collaborate with program chairs at major ML conferences to find solutions to the AI/ML research cycle outpacing current conference cycles, making many works outdated by the time of conferences. Second, we propose a future workshop to address the issue by informing the AI/ML community and gathering solutions from community members. Third, we call for a panel of influencers, industry, and academics to discuss how the evolving landscape of AI/ML research can be improved.

\section{Related Work}
\label{sec:related}

For over a decade now, social media platforms--namely $\mathbb{X}$--have been studied as an influential means of scholarly communication. \citet{Darling2013TheRO} discusses the potential role of Twitter in many stages of the 'life-cycle' of academic authorship and publication. \citet{Eysenbach2011CanTP} provides early evidence to support social media sharing as a predictor of higher citation count in medical publications. More recently, studies have shown a statistically significant relationship between Twitter presence and citation counts across fields, especially after the first tweet \cite{Peoples2016TwitterPC, Vaghjiani2021SocialMA}. Others suggest that the correlation between tweets and citations is insignificant through randomized trials \cite{Tonia2016IfIT, Branch2023ControlledEF}. 

We note that the AI/ML community is much more active on platforms like $\mathbb{X}$. While many previous works examine fields with a much lower volume of repository publications, the short research cycle and the need for quick dissemination of results, coupled with the rise of arXiv, motivates us to study the problem from the perspectives of modern AI/ML research lens. Additionally, the literature thus far has focused on social media as a forum for multi-user online discussion. In our work, we examine top-down dissemination from singular influencers, with many more followers than previously studied.  We also contribute a geographic and gender analysis of influencer-sharing patterns and provide recommendations to conferences, influencers, and the wider community on managing the evolving field of Artificial Intelligence research. 

\section{Data Collection}
\label{sec:data}

\begin{table}[t]
    \centering
    \begin{tabular}{l|c|c}
        \toprule
        & \textbf{AK} & \textbf{Komatsuzaki} \\
        \midrule
        \# Unique Papers & 9,171 & 1,273 \\
        \# with all attributes & 8,259 (90\%) & 1,191 (94\%) \\
        \# Matched Papers & 6,890 (75\%)& 955 (78\%) \\
        \bottomrule
    \end{tabular}
    \caption{The number of unique papers tweeted by each influencer, the subset of papers with all desired attributes available through S2 (Table~\ref{tab:attributes}), and the papers finally included in our analysis after matching. }
    \label{tab:target_set}
\end{table}

We model our analysis on retrospective cohort studies, in which a treatment and control group with identical underlying covariates are compared to determine the average treatment effect. In our case, we assume that a paper's citation count is most strongly influenced by elapsed time, quality, topic, and author prominence. While elapsed time is simple to measure, paper quality and topic are difficult to quantify. We use the publication venue and year as a proxy for quality and use a text embedding of the paper's title and abstract to approximate the topic. We use author citations and h-indices to adjust for their prominence.

As such, our data collection process consists of three parts: (1) collecting the Target Set, the papers tweeted by @\_akhaliq and @arankomatsuzaki, (2) collecting a large dataset of potential papers to match against, and (3) forming the Control Set by matching papers from (1) to papers from (2) with respect to the year of publication, the publication venue, and a text embedding of title and abstract. We detail each step below.

\begin{table}[tbp]
    \centering
    \begin{tabular}{c|c|c|c}
        \toprule
        \multicolumn{2}{c}{\textbf{AK}} & \multicolumn{2}{|c}{\textbf{Komatsuzaki}} \\
        \midrule
        Name & Freq. & Name & Freq.  \\
        \midrule
        S. Levine & 85 & L. Zettlemoyer & 31 \\
        Furu Wei & 82 & Quoc V. Le & 28	\\
        Jianfeng Gao & 71 & Yi Tay & 26 \\
        L. Zettlemoyer & 64 & Furu Wei & 23 \\
        Ziwei Liu & 62 & M. Dehghani & 18  \\
        \bottomrule
    \end{tabular}
    \caption{The top 5 most common authors shared by each user and the number of papers where they are credited. }
    \label{tab:top_authors}
\end{table}

\noindent \textbf{Target Set. }
The Target Set is collected by searching for document identifiers in the $\mathbb{X}$ feeds of both influencers (identifiers are found in links to \url{arXiv.org} or \url{huggingface.co/papers}). We use the Semantic Scholar (S2) API to query every document for desired attributes (Table~\ref{tab:attributes}). Notably, some papers are either not available on S2 or are not freely accessible. Hence, we remove any paper that lacks any required attribute. The number of unique papers and papers with all attributes for each influencer is shown in Table \ref{tab:target_set}.

We find that roughly 90\% of papers tweeted by either influencer have all attributes available. While we express moderate caution about the excluded minority of papers, as they may be systematically different from those included, we believe the many papers remaining in our analysis are sufficient to draw meaningful conclusions.

Additionally, we find that half of the papers are tweeted within 24 hours of their initial ArXiV release, with 95\% of tweets occurring within a week. This verifies that our influencers do indeed tweet freshly posted papers. We also find significant overlap in sharing, with 65\% of Komatsuzaki-shared papers also tweeted by AK.

\begin{table}[t]
    \centering
    \begin{tabular}{l}
        \toprule
        \texttt{title, abstract, year, venue,} \\
        \texttt{citationCount, embedding.specter\_v2,} \\ 
        \texttt{authors.\{name, affiliations, } \\ 
        \texttt{hIndex, citationCount\}} \\
        \bottomrule
    \end{tabular}
    \caption{The desired attributes for each paper, queried from S2.}
    \label{tab:attributes}
    
    \vspace{2em}
    \begin{tabular}{l}
        \toprule
        \texttt{year, venue, open\_access, n\_authors,} \\
        \texttt{first(h-index), max(h-index),} \\ 
        \texttt{first(citations), max(citations),} \\ 
        \texttt{topic\_embedding} \\
        \bottomrule
    \end{tabular}
    \caption{The covariates used for control matching. \texttt{first} = the corresponding value for the paper's first author, \texttt{max} = the maximum corresponding value among a paper's authors.}
    \label{tab:covariates}
\end{table}

\noindent \textbf{Control Set. }
To build our Control Set, we first collect a corpus of papers presented at the same venues and in the same years as those in the Target Set. Specifically, for every instance of a paper published in year $y$ at venue $v$, we query S2 for all papers published in year $y$ at venue $v$. 

We follow the recommendations in \cite{King2019} by performing exact matching on our categorical (year, venue, open\_access\footnote{Due to the nature of influencer sharing, all sampled papers are open access.}) and binned variables (n\_authors, h-index, author cites), and Euclidean distance\footnote{Note that Euclidean distance is equivalent to cosine similarity for normalized vectors.} matching on our continuous variables (topic\_embedding). Optimal matching can be reduced to the linear sum assignment problem, for which many efficient algorithms exist \cite{Crouse2016matching}. We use the implementation available in SciPy \cite{2020SciPy-NMeth}.

Using this methodology, we match each paper in the Target Set to a paper in the Control Set with respect to the gathered covariates (Table~\ref{tab:covariates}). We exclude any paper for which we cannot find an exact match on categorical variables. For matching on topic embeddings, we use SPECTER2 \cite{Singh2022SciRepEvalAM}. Matches between papers were strong, and the distribution of cosine similarities and example matched pairs are detailed in Appendix~\ref{app:matching}.

\begin{figure}[tbp]
    \centering
    \includegraphics[width=.9\linewidth]{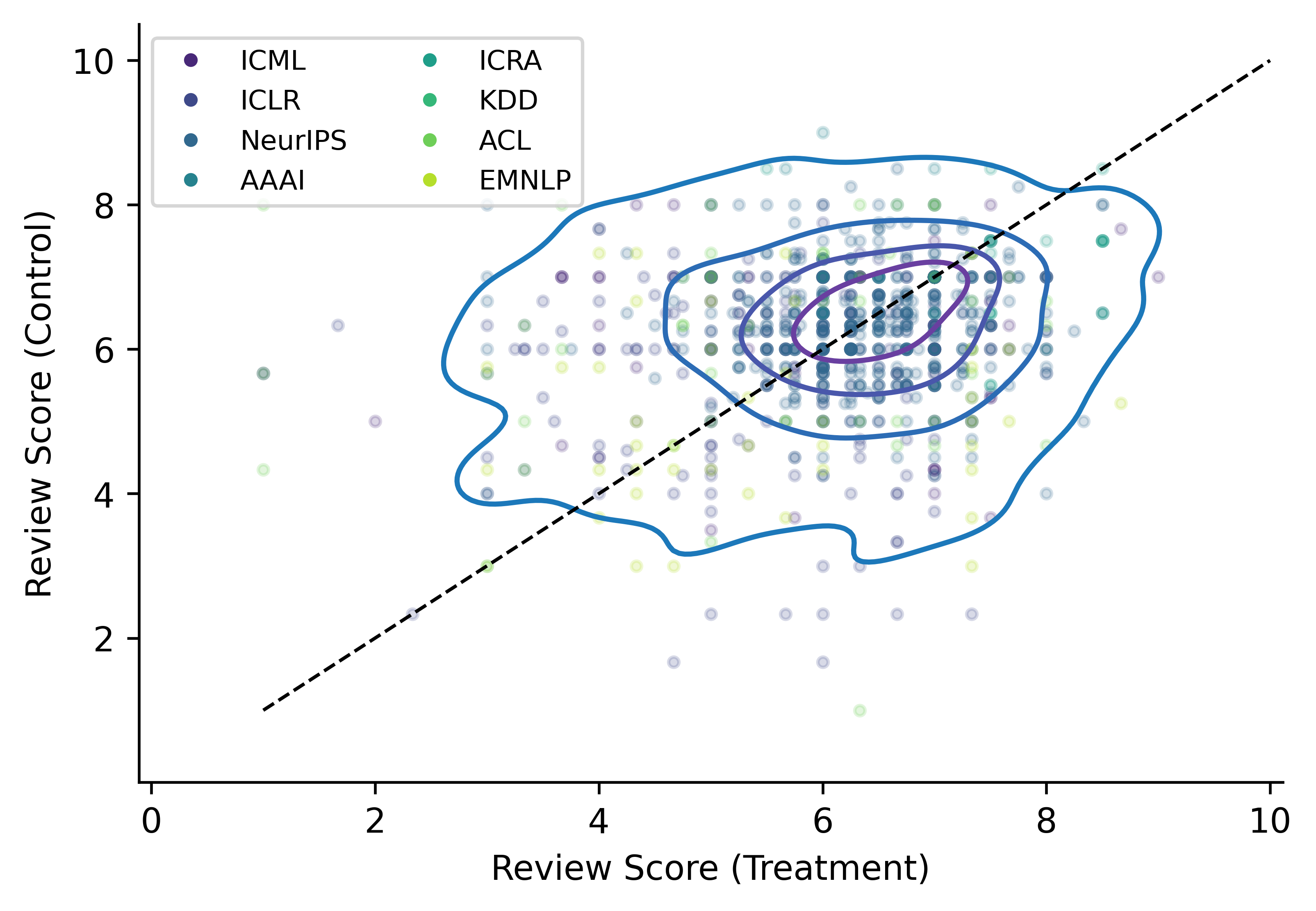}
    \caption{Mean OpenReview scores of tweeted papers vs non-tweeted controls from 8 major ML conferences, the kernel density estimate of the joint distribution of scores, and the identity line (dotted). This shows the quality of papers in both sets are more or less equivalent.}
    \label{fig:review-scores}
\end{figure}

\noindent \textbf{Review Scores. }
To determine if we have successfully controlled for quality through our matching, we will look at the review scores of experimental and control pairs from some selected conferences. We extract the review data using \citet{OpenReviewAPI}. Across both paired datasets, we found 939 out of 7222 unique pairs with available scores. 

We plot each treatment paper's mean review score against its paired control's mean score (Figure~\ref{fig:review-scores}). For conferences that do not use numbered review scores, we assign numbers based on those of other conferences (7: Accept, 5: Borderline Accept, 3: Borderline Reject, 2: Reject). 

Using the same three significance tests from Table~\ref{tab:2-sample_tests}, we do not find sufficient evidence to reject the null hypothesis that the control and experimental scores are from the same distribution ($p$-value $> 0.2$). Assuming that mean review scores are an accurate measure of paper quality, we conclude that we have effectively controlled for paper quality in our matching.

\begin{figure*}[!t]
    \centering
    \begin{subfigure}[b]{0.49\textwidth}
        \centering
        \includegraphics[width=\textwidth]{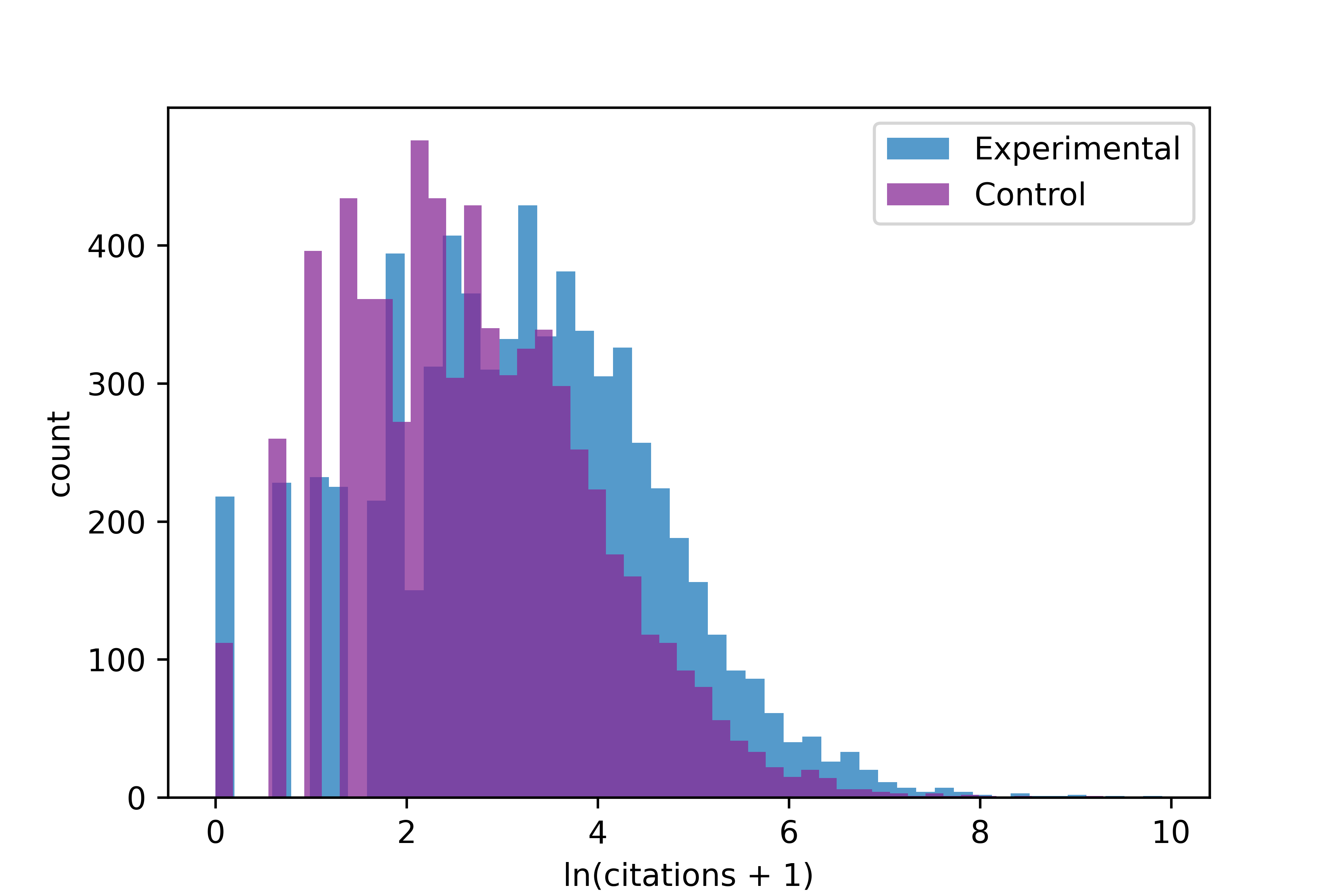}
        \caption{@\_akhaliq Dataset Histogram}
        \label{fig:hist:akhf}
    \end{subfigure}
    \hfill
    \begin{subfigure}[b]{0.49\textwidth}
        \centering        
        \includegraphics[width=\textwidth]{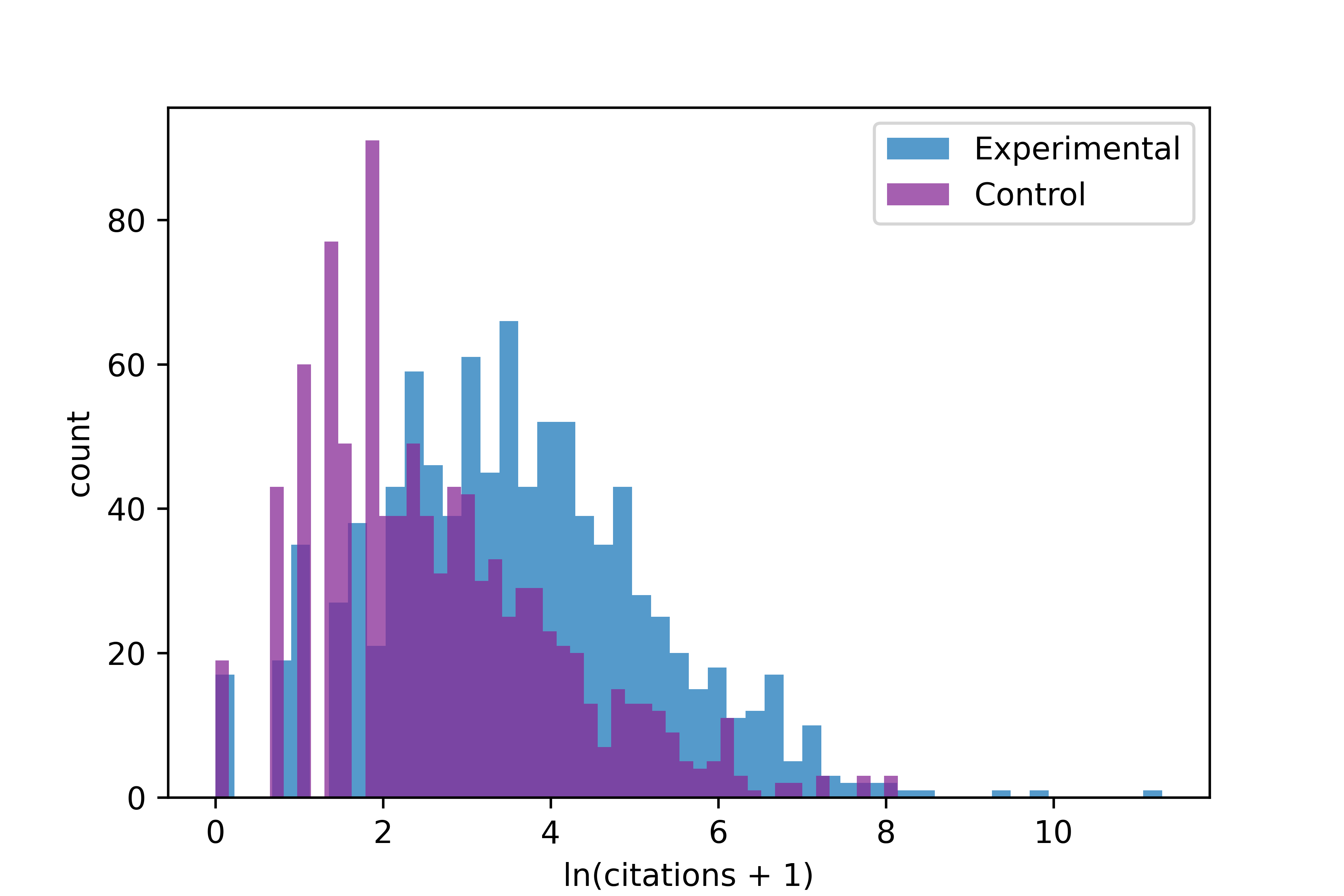}
        \caption{@arankomatsuzaki Dataset Histogram}
        \label{fig:hist:akgt}
    \end{subfigure}
    \begin{subfigure}[b]{0.49\textwidth}
        \centering
        \includegraphics[width=\textwidth]{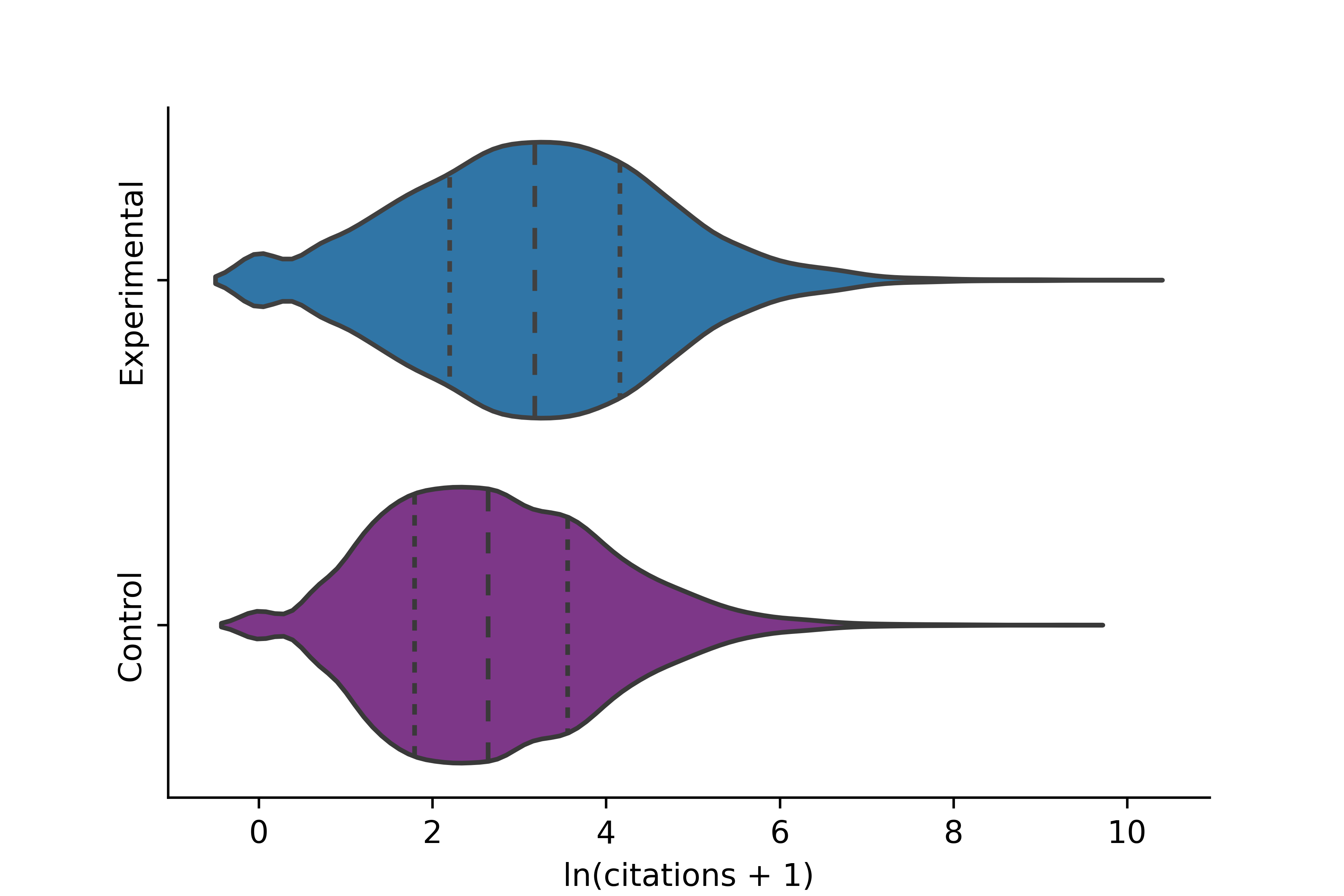}
        \caption{@\_akhaliq  Dataset Violin Plots}
        \label{fig:violin:akhf}
    \end{subfigure}
    \hfill
    \begin{subfigure}[b]{0.49\textwidth}
        \centering
        \includegraphics[width=\textwidth]{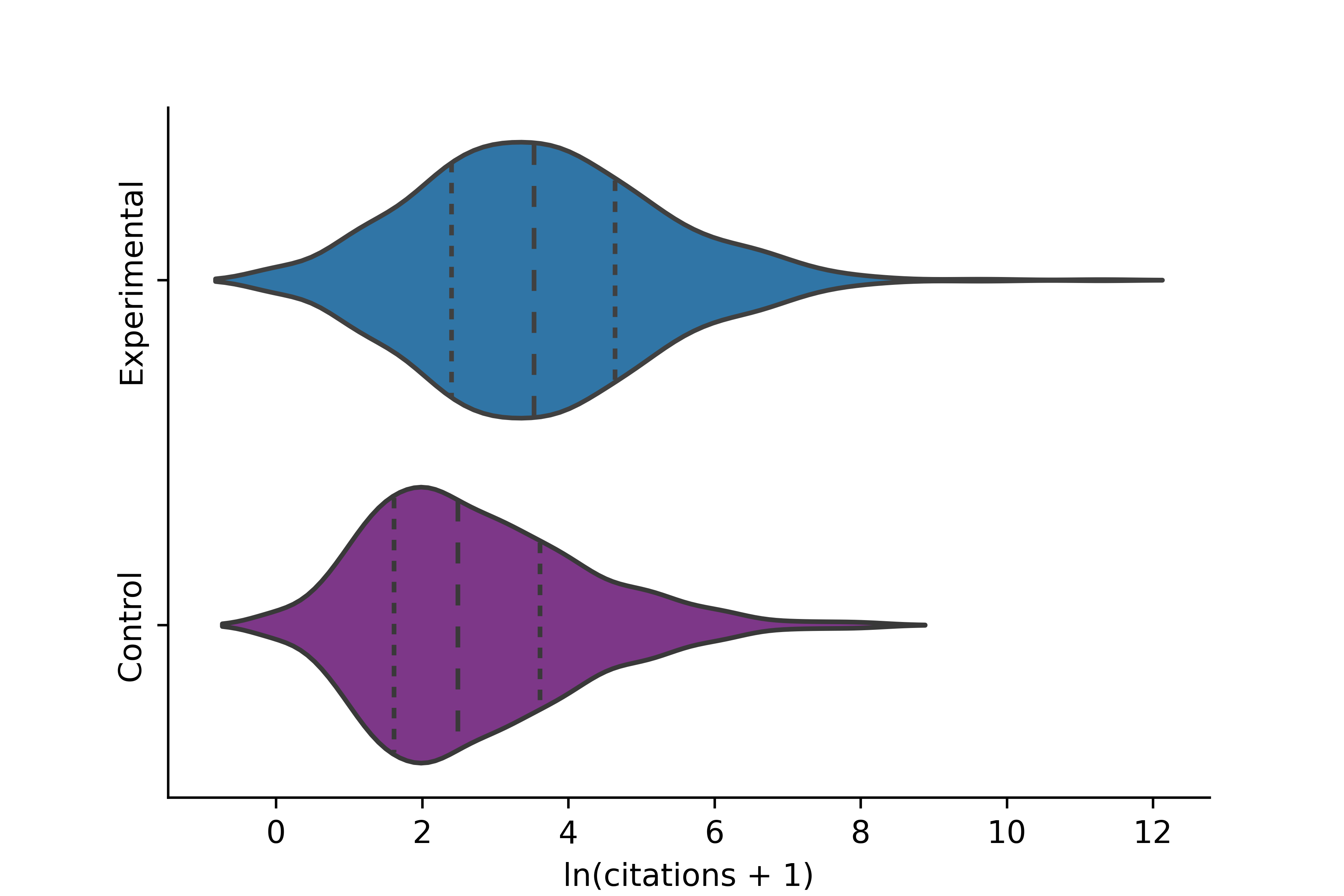}
        \caption{@arankomatsuzaki Dataset Violin Plots}
        \label{fig:violin:akgt}
    \end{subfigure}
    \caption{Plots showing the distribution of citations in the two experimental datasets and matched control samples. Citation counts are scaled with the natural logarithm using \texttt{numpy.log1p}. Both comparisons show that papers shared by influencers have attained significantly higher citations for all three quartiles than those in the control sets.}
    \label{fig:plots}
\end{figure*}

\section{Analysis}
\label{sec:analysis}

To answer the central question of our work, we compare the impacts of papers shared by AK and Komatsuzaki against our control set. Then, we conduct multivariate analyses by geographic distributions and author attributes in their selected papers.

\begin{figure*}[!t]
    \centering
    \begin{subfigure}[b]{0.49\textwidth}
        \centering
        \includegraphics[width=\textwidth]{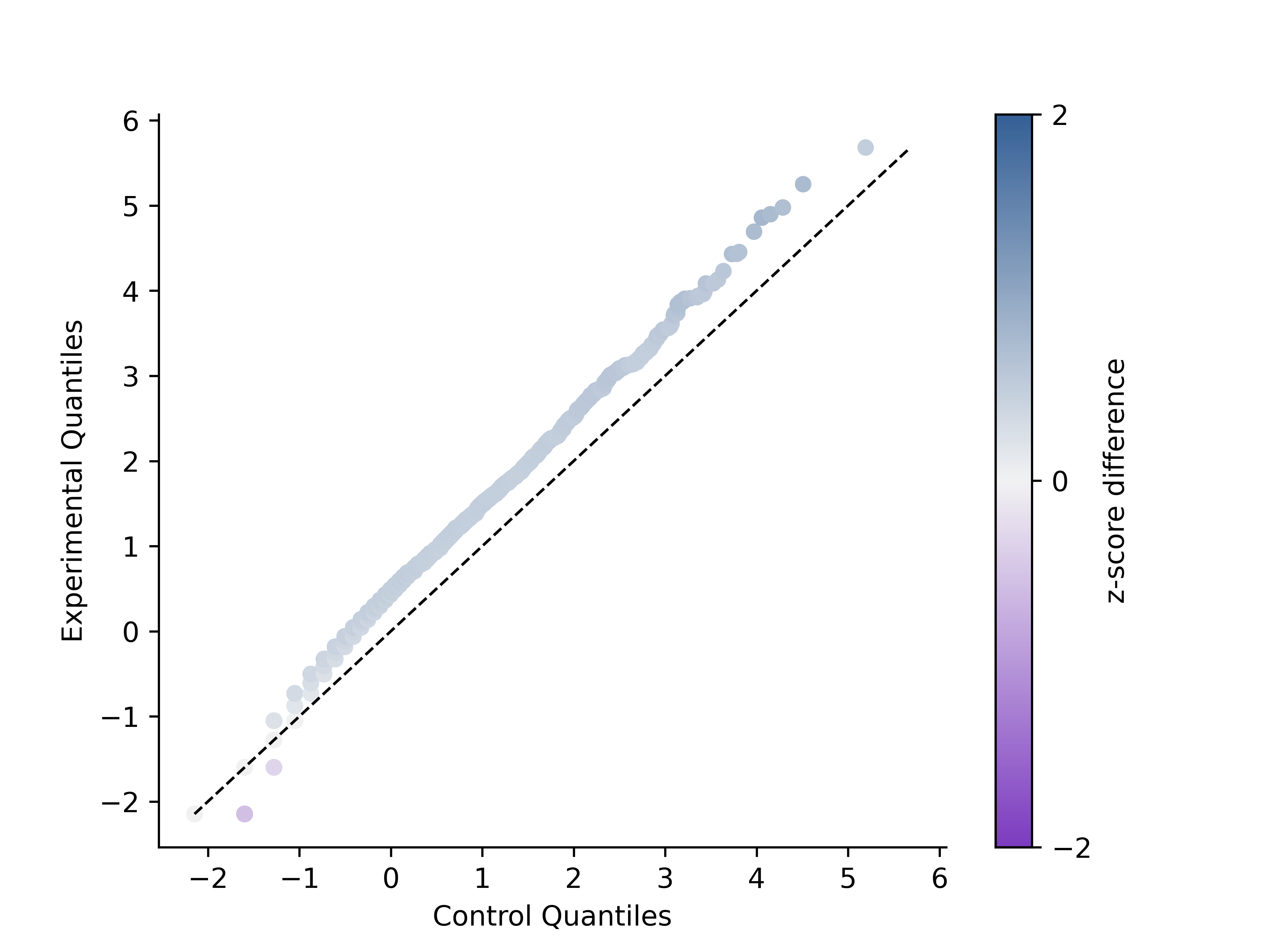}
        \caption{@\_akhaliq Dataset}
        \label{fig:qq:akhf}
    \end{subfigure}
    \hfill
    \begin{subfigure}[b]{0.49\textwidth}
        \centering
        \includegraphics[width=\textwidth]{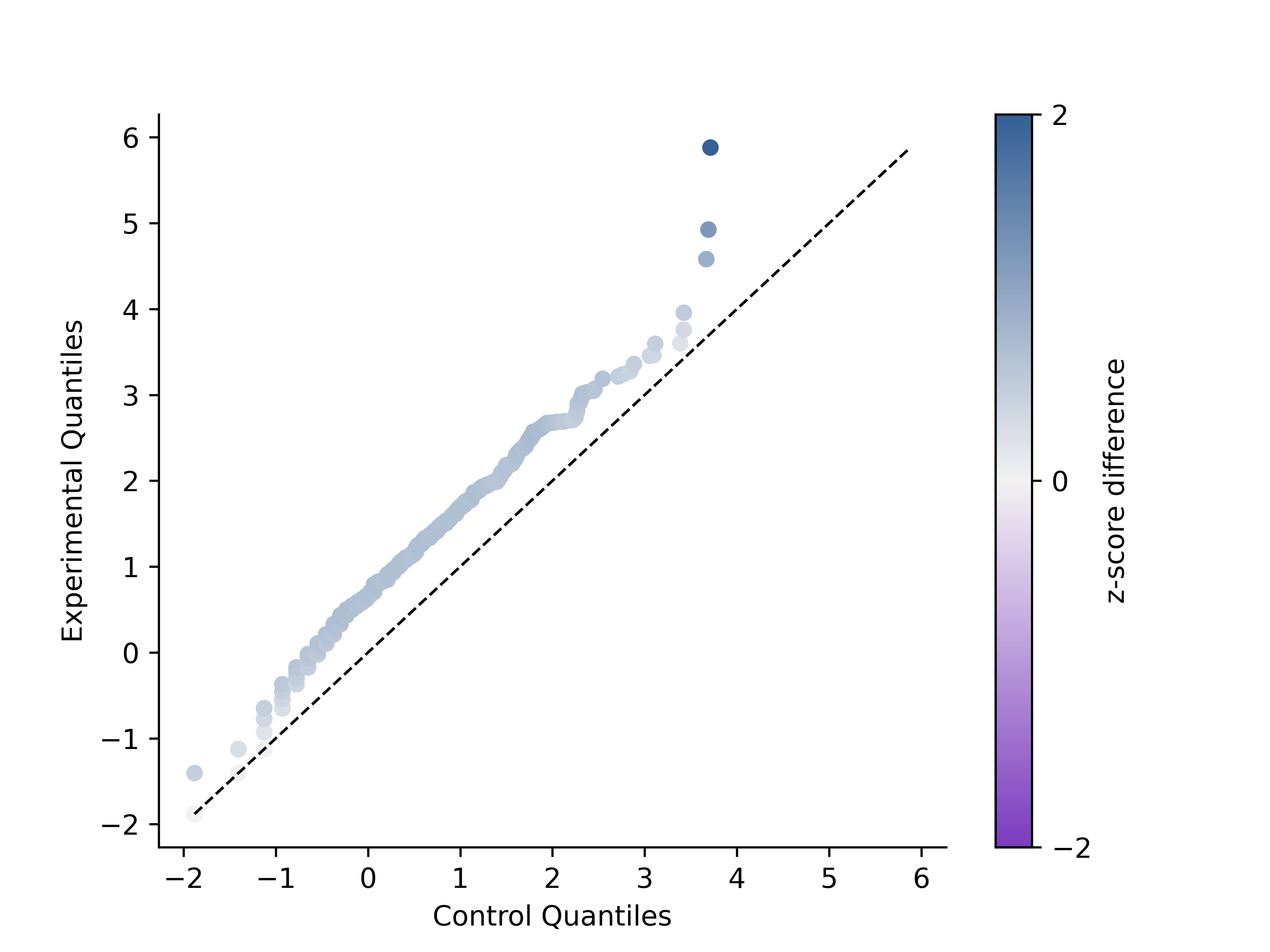}
        \caption{@arankomatsuzaki Dataset}
        \label{fig:qq:akgt}
    \end{subfigure}
    \caption{2-Sample Q-Q Plots comparing the experiment and control set distributions across every quantile. To build the plot, citation counts are log-scaled, normalized to the control distribution (z-scores), sorted, and paired in order. The dotted line shows an equal distribution; any points above the line show a higher experimental quantile, and vice versa. The plots show that both experimental distributions are consistently higher, especially closer to the median.}
    \label{fig:qq}
\end{figure*}

\noindent \textbf{Contrasting Analysis. }
For the contrasting analysis, we will test the following hypotheses for correlation:
\vspace{-0.5em}
\begin{itemize}
    \setlength\itemsep{-0.4em}
    \item \textbf{Null:} Influencer-shared papers have the same citation count as others in the same field.
    \item \textbf{Alternative:} Influencer-shared papers have a higher citation count than others in the same field.
\end{itemize}
\vspace{-0.5em}

\begin{table}[bp]
    \centering
    \begin{tabular}{l|c|c}
    \toprule
    \textbf{Test} & \textbf{AK} & \textbf{Komatsuzaki} \\
    \midrule
    Epps-Singleton & $p<0.0001$ & $p<0.0001$ \\
    Kolmogorov-Smirnov & $p<0.0001$ & $p<0.0001$ \\
    Mann-Whitney U & $p<0.0001$ & $p<0.0001$ \\
    \bottomrule
    \end{tabular}
    \caption{2-sample Statistical significance tests for the difference in distributions of the two experiment and control datasets. These tests suggest a statistically significant difference in the experimental and control distributions for both influencer datasets.}
    \label{tab:2-sample_tests}
\end{table}

We compare our paired target and control sets, as described in Section~\ref{sec:data}. We find that papers tweeted by AK have a median citation count of 24 (95\% CI: 23, 25) versus 14 (95\% CI: 13, 15) in the control group, and papers tweeted by Komatsuzaki have a median citation count of 31 (95\% CI: 27, 34) versus 12 (95 \% CI: 10.5, 13.5). Visually, we can see that both experimental set distributions are skewed toward higher citation counts when compared to their corresponding control sets (Figure~\ref{fig:plots}). In the violin plots (Figures~\ref{fig:violin:akhf} and~\ref{fig:violin:akgt}), the three quartiles and max values are all higher in both of the shared paper distributions compared to the controls. In the 2-Sample Q-Q plots (Figure~\ref{fig:qq}), we can see that the normalized quantiles are consistently higher for the test distributions.

Finally, we establish statistical significance with three tests comparing the distributions of the experimental data with that of the control sets, Epps-Singleton (ES), Kolmogorov-Smirnov (KS), and Mann-Whitney U (MWU), none of which assume normal distribution, which is essential for our data. Table~\ref{tab:2-sample_tests} shows the results, all with $p$-values well below even a stringent $\alpha = 0.001$. From this, we can strongly reject the null hypothesis that the citation distributions for the influencer-shared papers and the control papers are the same.

Overall, the correlation between influencer tweets and citation count--and not review scores--points to a shift in how the community finds and reads papers. While traditionally, top conference acceptance (i.e. review score) has been a primary indicator of future citation count \cite{Lee2018PredictivePO}, we have shown that the sharing practices of far-reaching influencers are now a significant indicator of future research impact through citations. 

\begin{figure*}[!t]
    \centering
    \includegraphics[width=0.75\textwidth]{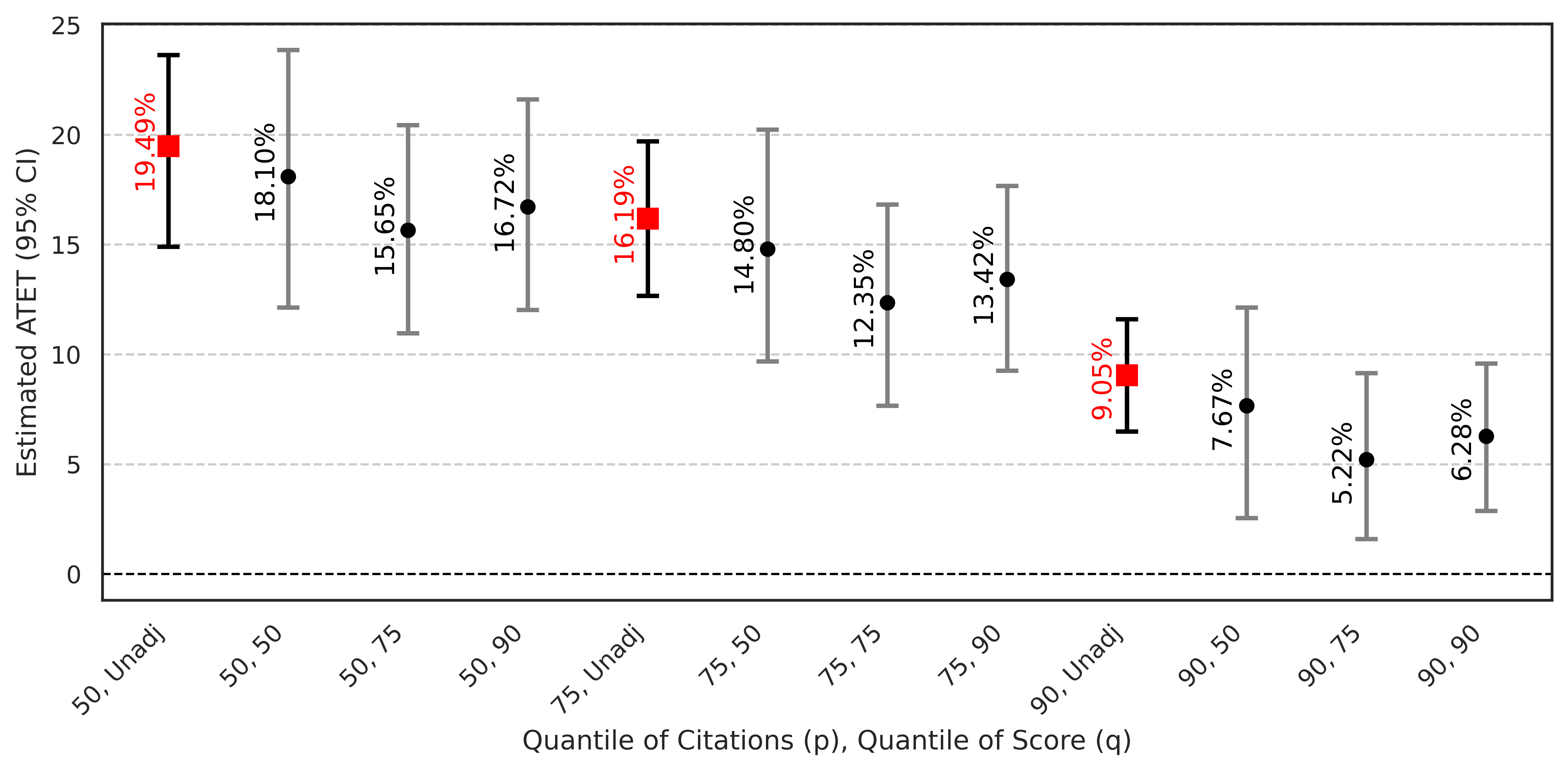}
    \caption{Forest plot of ATET and confidence intervals approximated with negative outcome control (NOC). Larger positive values indicate a stronger positive causal effect of the treatment (influencer sharing) on the outcome (a paper being "highly-cited"). To ensure robust results, we vary the quantile threshold for "highly-cited" and "highly-scored" papers; "Unadj." values (red squares) show the effect estimate before applying NOC. These results, where no confidence intervals contain 0\%, indicate a significant positive causal effect of influencer sharing on paper citations. }
    \label{fig:atet}

\end{figure*}

\noindent \textbf{Causal Inference: Setup. }
Although we have shown a significant correlation between influencer sharing and citations, we have not investigated a causal link. For this, we turn to the model, techniques, and assumptions presented in \citet{Elazar2023EstimatingTC}. For additional background and details, see Appendix~\ref{app:causalinf} 

We aim to estimate the causal effect of the treatment $A$, indicating that the paper was shared by influencers\footnote{To increase sample size and simplify our analysis, we combine our test sets for this portion of the analysis.}, on the outcome $Y$, if the paper is "highly-cited" or "less-cited." Besides $A$, there are a number of other factors, known as \textit{confounders}, that can affect $Y$. We divide these confounders into two sets: \textit{observed} $C$--which are the same used in our matching (Section~\ref{sec:data})--and \textit{unobserved} $U$--which are difficult to quantify or measure (e.g. novelty, contribution, hype). 

To debias the effects of unobserved confounders $U$, we employ the negative outcome control (NOC) framework \cite{Card1993MinimumWA, Lipsitch2010NegativeCA}. By finding a \textit{negative control outcome} $N$ that shares the same confounders as $Y$, but is not causally affected by $A$, we can attempt to correct for the bias introduced by $U$. To this end, we select a paper's average review score for $N$, which we have already shown is not significantly correlated to $A$ (Figure~\ref{fig:review-scores}) and which we believe share many of the same confounders (e.g., quality, contribution, author experience). However, another NCO can be substituted if deemed more suitable.

To account for fine-grained citation counts and review scores in comparison to the binary nature of $Y$ and $N$, we define $Y_p$ and $N_q$. For both, their value is 1 if the corresponding real-valued quality (citation count or average review score) is above the $p^\text{th}$ (or $q^\text{th}$) quantile of the relevant sample distribution. We will use values of $\{50,75,90\}$ for both $p$ and $q$ in our analysis.

Finally, we estimate the causal effect of influencer sharing on paper citations with the \textit{average treatment effect of the treated}. Using the difference-in-difference assumption with NOC, we can write our estimate \cite{Elazar2023EstimatingTC}:
$$\text{ATET} = \mathbb{E}[Y_{A=1}-N_{A=1}] - \mathbb{E}[Y_{A=0}-N_{A=0}]$$
where ATET $\in[-100\%,100\%]$ values closer to $0\%$ indicate weak to no effect of the treatment, and large absolute values indicate a stronger effect.

\noindent \textbf{Causal Inference: Results. }
First, we estimate the effect of $A$ on $Y$ without accounting for the effects of $U$. We estimate ATET (95\%-bootstrapping CI) to be 19\% (15-24), 16\% (13-20), 9\% (6-12) for $p=\{50,75,90\}$, respectively. This confirms the significant association between influencer sharing and citations we investigated during the contrasting analysis. However, this does not preclude the existence of strong confounders.

For that, we will use NOC to attempt to debias the effects of unobserved confounders, including quality. We record our results in Figure~\ref{fig:atet} and observe the following: (1) debiased effects are smaller than the unadjusted estimates, indicating our NCO has succeeded in accounting for some unobserved confounders; (2) all of our estimated effects are significant to the 95\% level, because none contain 0 in their confidence interval; (3) influencer sharing is most successful at increasing the citations in lower percentiles--papers with very high citation counts experience a smaller (yet still significant) effect compared to those with fewer citations.

\begin{figure*}[ht]
    \centering
    \begin{subfigure}{.49\textwidth}
        \centering
        \includegraphics[width=\textwidth]{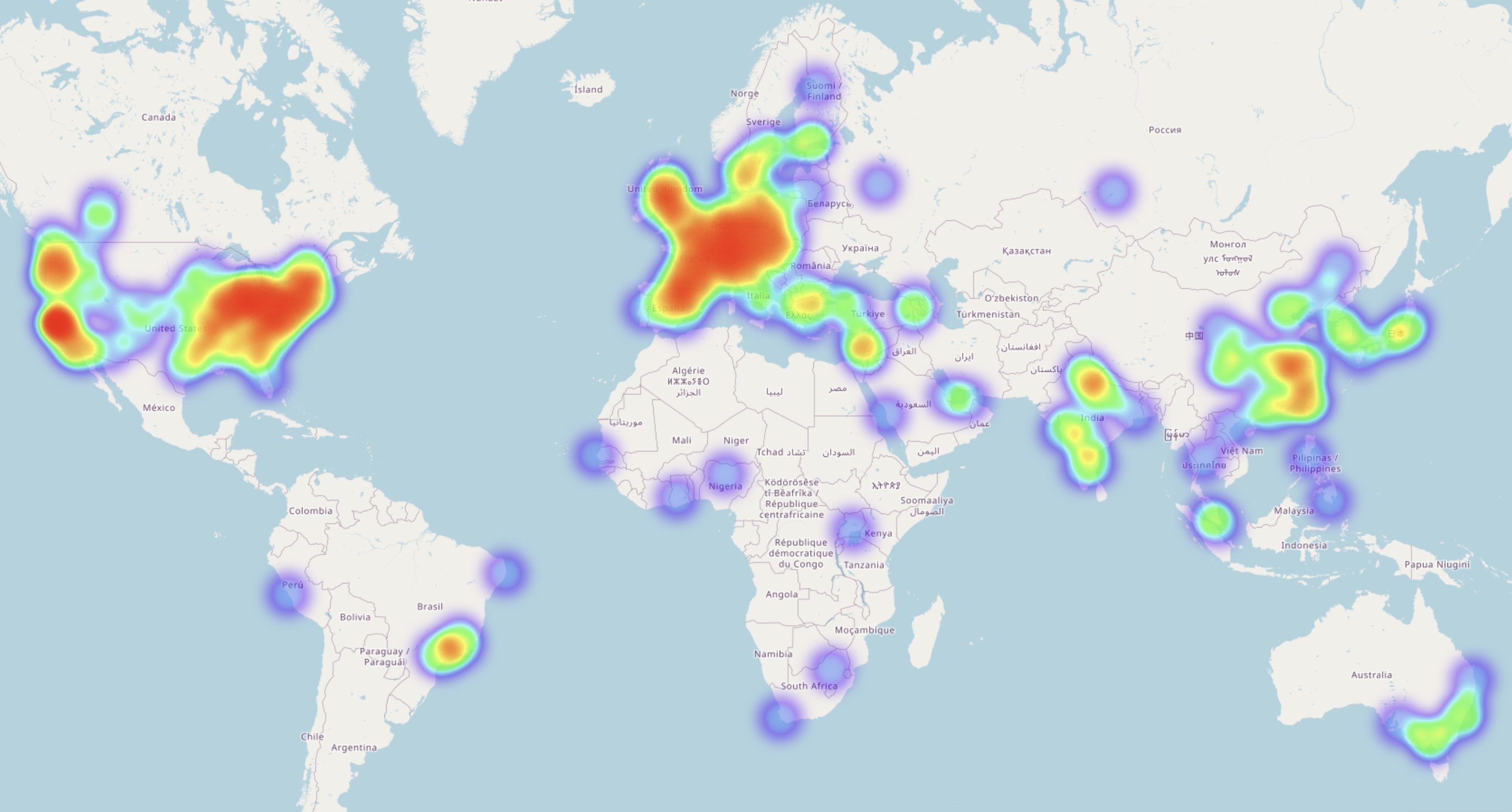}
        \caption{affiliation heatmap of @\_akhaliq-shared authors}
        \label{fig:geo-map:akhf}
    \end{subfigure}
    \hfill
    \begin{subfigure}{.49\textwidth}
        \centering
        \includegraphics[width=\textwidth]{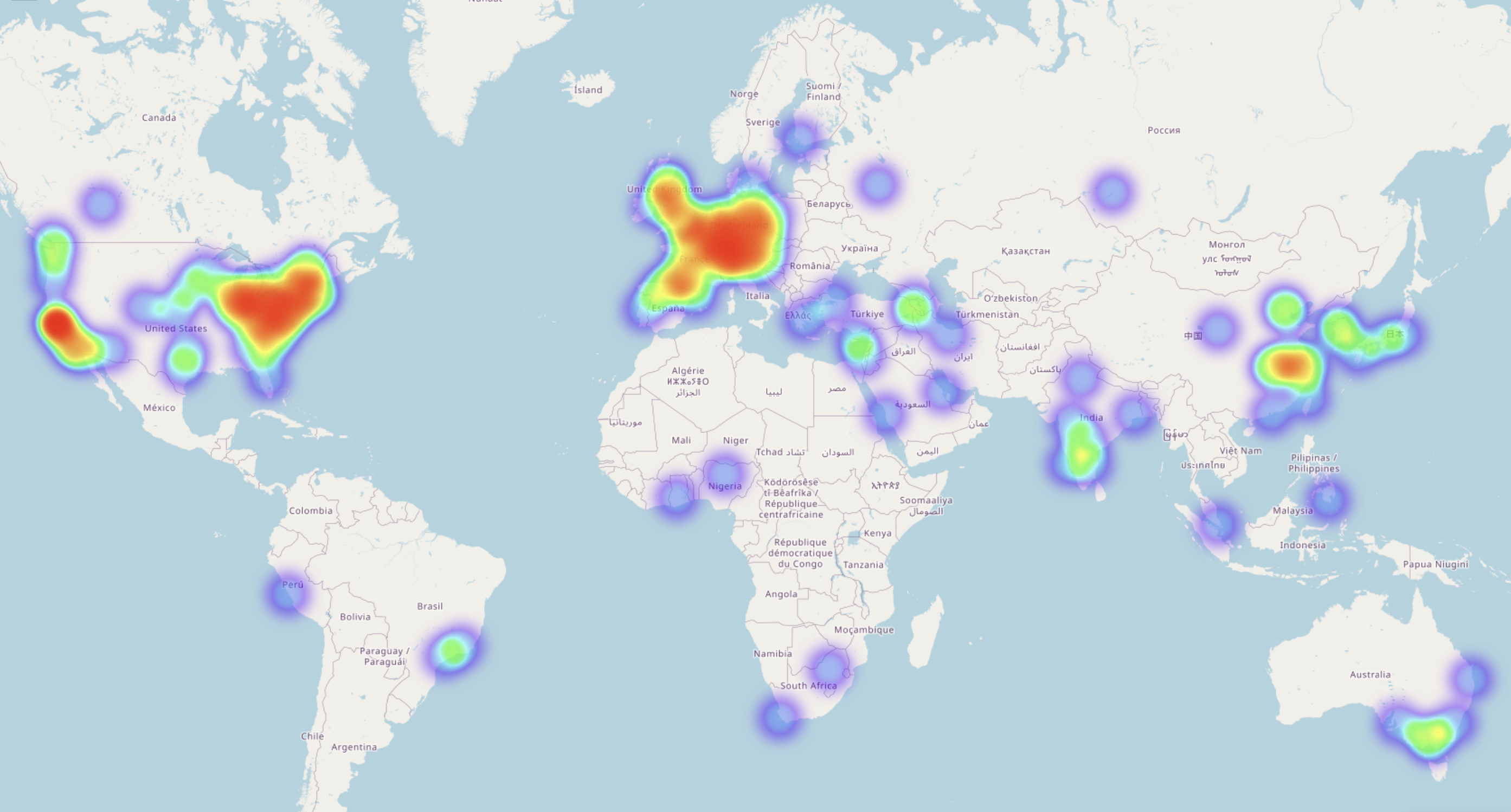}
        \caption{affiliation heatmap of @arankomatsuzaki-shared authors}
        \label{fig:geo-map:akgt}
    \end{subfigure}
    \caption{Geographic heatmaps of the unique affiliations of influencer-shared paper authors. High density (red) on the map represents a large number of unique institutions and not a large number of papers, citations, or authors from the given area. Both influencers shared from institutions around the world, with especially large hotspots in the US and Europe.}
    \label{fig:geo-maps}
\end{figure*}

\noindent \textbf{Geographic Distributions. }
In exploring the evolving landscape of machine learning (ML) paper dissemination, it is essential to consider the implications of a more centralized curation model, particularly as it relates to geographic and gender diversity in scholarly works. Our approach is to present data and observations that highlight trends \textit{\textbf{and do not attribute intentional bias to the influencers involved}}.

\begin{figure}[htbp]
    \centering
    \includegraphics[width=\linewidth]{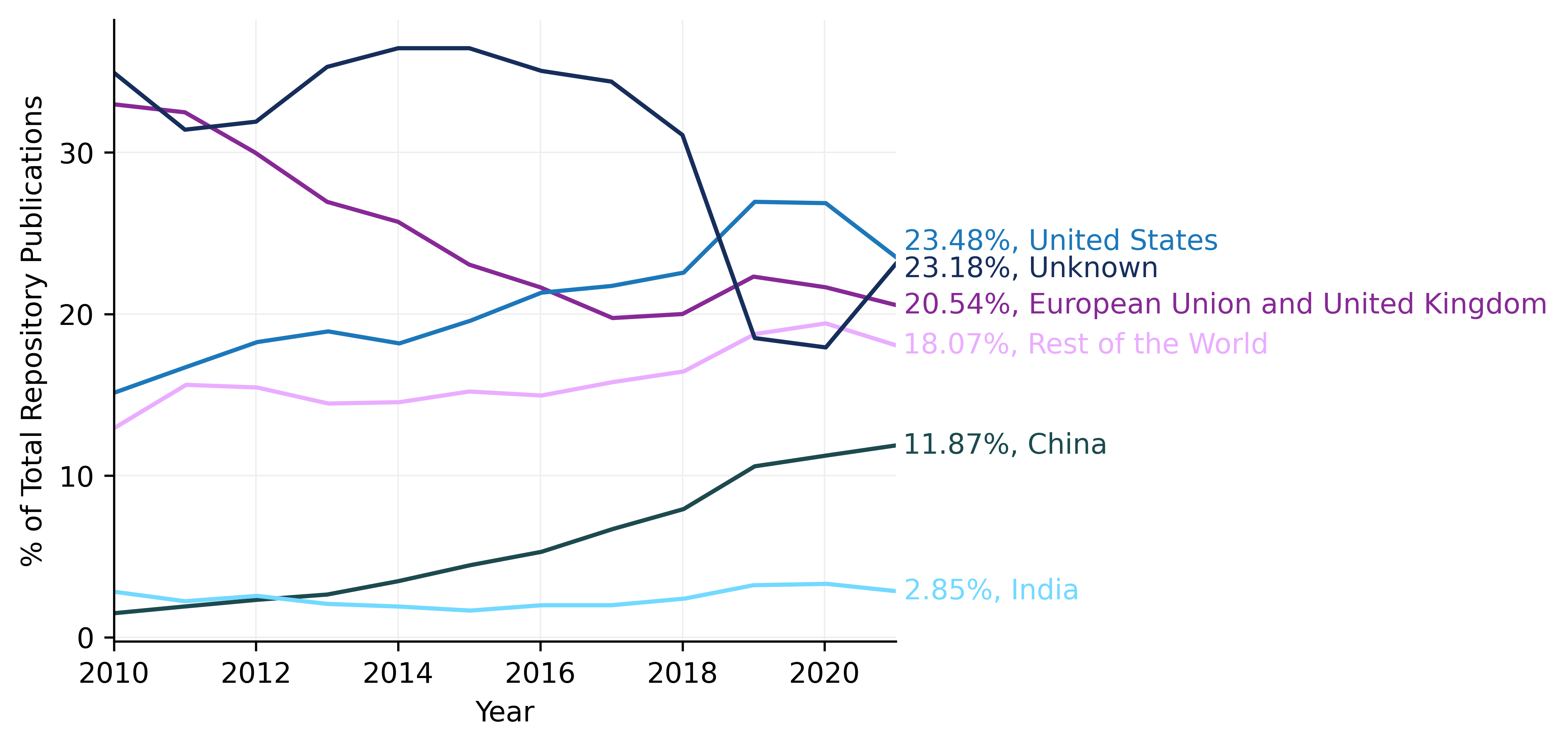}
    \caption{AI repository publications by geographic area and publication year, 2010–2021. Data extracted from Figure 1.1.19 of the Stanford HAI 2023 AI Index Report \cite{maslej2023artificial}.}
    \label{fig:geo-plots:hai}
\end{figure}

Our analysis begins by examining the geographic distribution in the dissemination of ML papers. Given the American affiliations of AK and Aran Komatsuzaki, we explore whether this translates into a geographic skew in the papers they share. To contextualize our findings, we refer to the geographic distribution of AI repository publications from the Stanford HAI 2023 AI Index Report (Figure~\ref{fig:geo-plots:hai}). We choose to view this data in particular, as our selected influencers share papers from repositories (i.e. ArXiV).

\begin{figure*}[ht]
    \centering
    \begin{subfigure}{.49\textwidth}
        \centering
        \includegraphics[width=\textwidth]{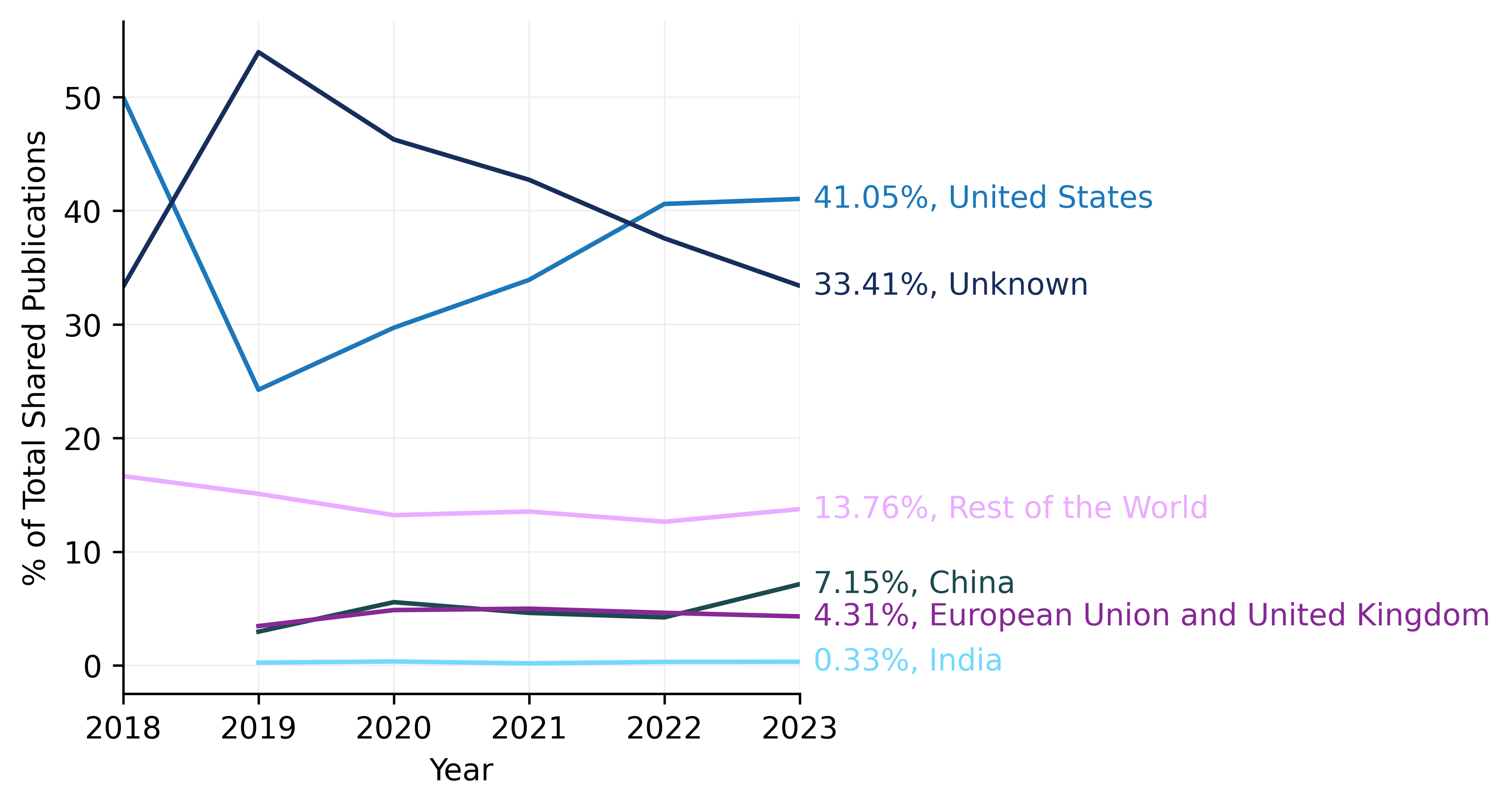}
        \caption{@\_akhaliq-shared papers by geographic area per year}
        \label{fig:geo-plots:test:akhf}
    \end{subfigure}
    \hfill
    \begin{subfigure}{.49\textwidth}
        \centering
        \includegraphics[width=\textwidth]{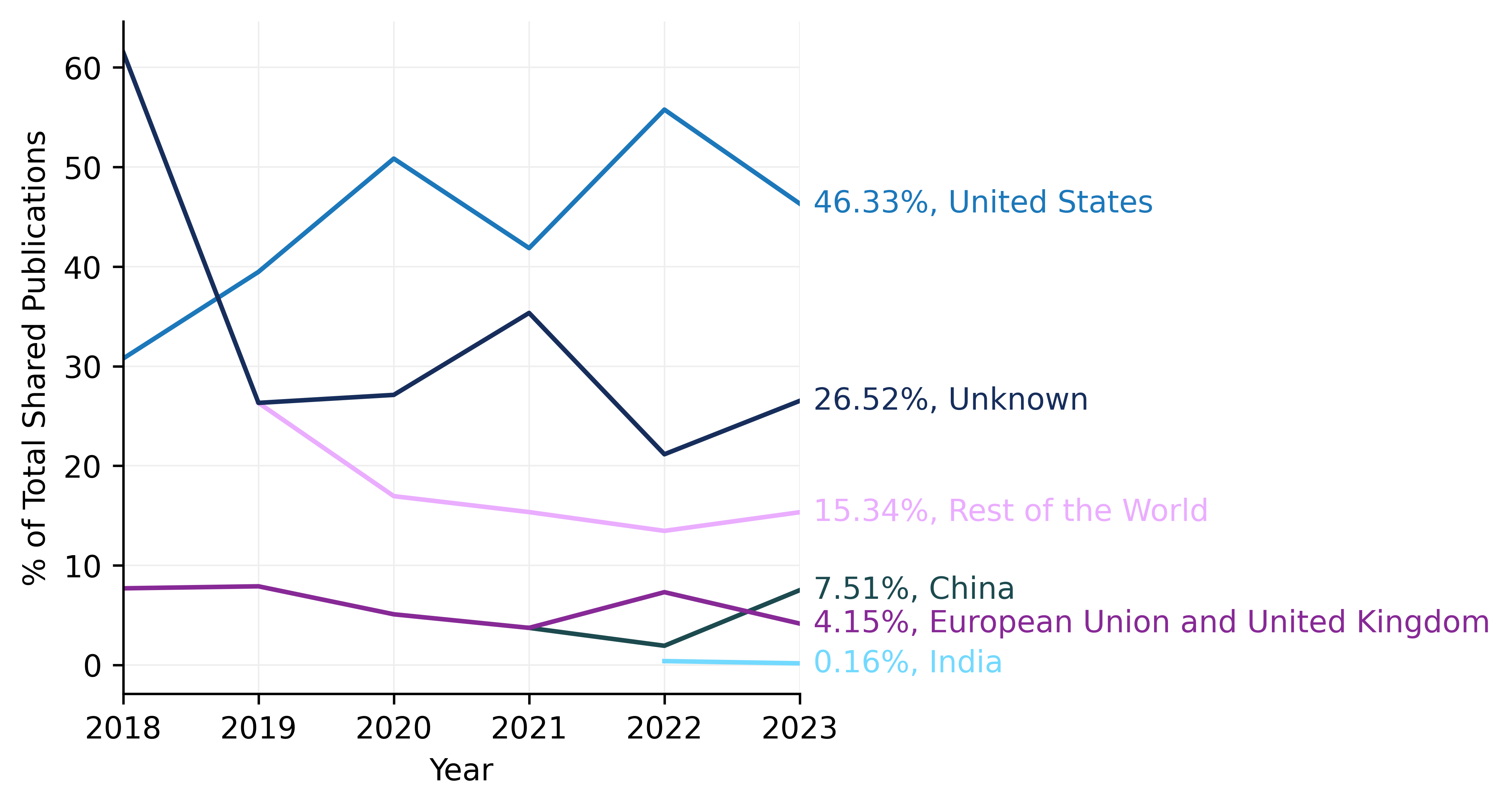}
        \caption{@arankomatsuzaki-shared papers by geographic area per year}
        \label{fig:geo-plots:test:akgt}
    \end{subfigure}
    \caption{Influencer-shared publications by geographic area and publication year, 2018-2023, from Semantic Scholar and DBLP affiliation data.}
    \label{fig:geo-plots:test}
\end{figure*}

To test this, we first collect the geographic data of the shared papers. First, we use S2 and \citet{dblp2023snapshot} to collect the affiliation data of all listed authors from each test set. We then utilize the Nominatim geocoding API to find the approximate latitude and longitude of each affiliation, manually adjusting visibly inaccurate coordinates. From this information, we find the country of each affiliation and then use majority voting to assign each publication a geographic area. At this point, we can see that both influencers share papers from around the world in Figure~\ref{fig:geo-maps}. Finally, we aggregate these countries into the same geographic areas used in the HAI Report and plot using a similar format (Figure \ref{fig:geo-plots:test}).

To account for the discrepancy between date ranges in the HAI Report and our influencers' activity, we will limit our analysis to the overlap between them. Additionally, we will focus on the range from 2018 to 2021, because of the low sample size of shared papers pre-2018--only 5 in total.

During these years, Figure~\ref{fig:geo-plots:hai} indicates a slight decline in the United States' share of AI repository publications following its peak. Concurrently, the European Union and United Kingdom demonstrate a modest uptick in publications after a consistent decline from 2010-2017, while China's share continues to rise.

In contrast, the sharing patterns of the influencers from 2018 to 2021, shown in Figure~\ref{fig:geo-plots:test}, demonstrate a notable deviation from these global trends. Specifically, both influencers' data exhibits a consistent focus on U.S.-affiliated papers, with a much smaller, relatively constant portion from other areas. The increase in U.S.-affiliated papers seems to be a change in reporting rather than a change in sharing practices. 

We must note, though, that using solely self-reported affiliations can have an inherent bias toward the United States. For example, many researchers affiliated with U.S.-based organizations are assigned to the United States despite working out of another area. Additionally, we must note the prominence of the "Unknown" category in the data of both influencers, where affiliations were not found. 

Nevertheless, our results highlight the potential for centralized individuals to shape the perceived narrative of AI research prominence.

\noindent \textbf{Gender Distributions. }
Beyond geographic diversity, gender diversity is crucial in Computer Science and Engineering, fields historically dominated by men. We extract author names and affiliations as described above. In this section, we filter only the first authors of each paper. For gender classification, we used the AMiner Scholar Gender Prediction API, which categorizes authors as "male," "female," or "UNKNOWN" based on name and affiliation--if available. The API uses a majority vote of the results from three sources: google image search and facial recognition, the Facebook Generated Names List \cite{Tang2011NamesGenderFacebook}, and WebGP \cite{Gu2016WebUP}. 

To ground our view of the overall gender distribution in the field, we reference the Taulbee survey's reported gender distribution of US Ph.D. awardees and faculty in CS and related fields from 2021-2022 \cite{Zweben_Bizot_2023}. To match the classifications we have available, we will consider the binary reported genders from the survey.

Our analysis revealed an 80:20 male-to-female ratio among authors with identifiable genders in the @\_akhaliq dataset and an 81:19 ratio in the @arankomatsuzaki dataset. These ratios align somewhat with the Taulbee survey's reported 77:23 ratio in computing Ph.D. awardees and deviated slightly more from the 76:24 ratio in faculty. These deviations may stem from a trend toward increasing female representation in the ML space; the survey is recent data, while our influencer data spans several years into the past.

\section{Discussion}
\label{sec:disc}

Our analysis suggests that ML influencers strongly affect paper visibility, indicating a change in how ideas are propagated through the community. We discuss the downstream implications of influencers on the community and make recommendations on how to help improve the paper curation problem and enhance equity in publication visibility.

\noindent \textbf{Influencers and the ML Community. }
Influencers serve as pivotal curators in the ML landscape. With the explosive growth of machine learning research, the community increasingly relies on social media to keep up-to-date with new developments. Their role in streamlining the dissemination process is akin to that of journalists in news media, making novel ideas and breakthroughs more accessible.

% However, we advise caution against excessive dependence on a limited set of information sources. Research inherently involves bottom-up exploration of a wide array of topics. Focusing on a handful of individuals' highlighted papers necessarily offers a narrow view of the research landscape. We encourage the community to keep the online academic space competitive, allowing for a diversity of highly visible ideas. This goal can be achieved through active participation in an open, community-driven curation process, enhancing the variety of prominent ideas. Additionally, we urge influencers to share a diverse array of ideas. This includes showcasing various techniques and subtopics within their areas of focus, thereby exposing community members to new approaches and concepts.

Yet, reliance on a select few narrows the full spectrum of research. We propose maintaining a competitive, open academic space with diverse, highly visible ideas, advocating for influencers to diversify showcased techniques and concepts. This encourages a broad exploration of the ML landscape, mitigating the risk of a homogeneous research narrative.

\noindent \textbf{Enhancing Equity. }
The move towards influencer-led dissemination opens paths to greater equity in the ML community. Our findings—pointing out the geographic and gender disparities in shared research—underscore the potential for a more inclusive academic dialogue. We clarify that our analysis does not suggest influencers currently exhibit bias; rather, it highlights an opportunity to mitigate existing inequities through proactive online engagement. By promoting diverse global perspectives and addressing gender disparities, influencers can foster a more equitable field. This effort aligns with our commitment to enhance equity without overemphasizing certain demographics, reflecting the concentration of ML research.

To stimulate community reflection and action, we recommend collaborating with conference chairs to address the pace of research, organizing workshops to explore dissemination improvements, and convening panels of influencers, industry, and academia to discuss enhancements in AI/ML research visibility. These steps aim to refine the research dissemination process and uplift the ICML community's engagement with emerging and diverse ideas.

\section{Conclusion}
\label{sec:concl}

Our study delves into the influence of social media influencers on the dissemination and recognition of academic work in AI/ML, specifically examining the impact of AK and Aran Komatsuzaki on platforms like $\mathbb{X}$ and Hugging Face. Our analysis demonstrates that papers endorsed by these influencers receive significantly more citations than those that are not, underscoring the role of influencers in not only amplifying the reach of specific research but also in shaping its visibility in the field. This influence extends beyond merely sharing higher-quality papers, highlighting their ability to promote substantial findings within the community.

We further discuss the dual role of influencers as both catalysts for visibility and curators of content, emphasizing the need for a balance in their influence to ensure a diversity of perspectives in the AI/ML research landscape. The implications of this influence are profound, suggesting a need for the academic community to reassess traditional methods of paper selection and review. Our findings encourage a dialogue among conference organizers and academic institutions to evolve the peer-review process and adapt to these changing norms. Additionally, we encourage future research into these trends in other scientific fields and the mechanisms behind the influence of social media on academic recognition.

%%%%%%%%%%%%%%%%%%%%%%%%%%%%%%%%%%%%%%%%%%%%%%%%%%%%%%%%%%%%%%%%%%%%%%%%%%%%%%%
%%%%%%%%%%%%%%%%%%%%%%%%%%%%%%%%%%%%%%%%%%%%%%%%%%%%%%%%%%%%%%%%%%%%%%%%%%%%%%%
% 8-page limit does not apply past this point
%%%%%%%%%%%%%%%%%%%%%%%%%%%%%%%%%%%%%%%%%%%%%%%%%%%%%%%%%%%%%%%%%%%%%%%%%%%%%%%
%%%%%%%%%%%%%%%%%%%%%%%%%%%%%%%%%%%%%%%%%%%%%%%%%%%%%%%%%%%%%%%%%%%%%%%%%%%%%%%

\section*{Impact Statement}
Recognizing the influential role of social media influencers in the academic community, particularly in the fields of AI/ML, it's important to gently encourage them to be aware of their impact. Influencers play a key role in shaping discussions and trends, and there is a growing need for them to consider the diversity and inclusivity of the research they share. This is not to undermine their contributions but to enhance the richness of academic discourse. Similarly, the academic community might benefit from reevaluating traditional metrics of research impact, embracing both peer-reviewed channels and digital platforms for a more holistic approach to recognizing scholarly work.

In doing so, we hope to spark future research into understanding any unintentional biases influencers might have, such as favoring research from prestigious institutions or specific regions. This is a subtle yet crucial aspect of ensuring a balanced representation in the academic landscape, preventing the overshadowing of significant but less publicized work. While influencers may have personal or commercial interests, we trust in their capacity to acknowledge and manage these to avoid conflicts of interest. This approach would help in preventing echo chambers and promoting a diverse range of perspectives in AI/ML research. Overall, we advocate for a collaborative and mindful effort among influencers and the academic community to foster an equitable and inclusive environment for scholarly discussions.

\section*{Acknowledgements}

This work was supported in part by the National Science Foundation Research Experience for Undergraduates under CAREER Award Grant No. 2048122.

\bibliography{custom}
\bibliographystyle{icml2024}

%%%%%%%%%%%%%%%%%%%%%%%%%%%%%%%%%%%%%%%%%%%%%%%%%%%%%%%%%%%%%%%%%%%%%%%%%%%%%%%
%%%%%%%%%%%%%%%%%%%%%%%%%%%%%%%%%%%%%%%%%%%%%%%%%%%%%%%%%%%%%%%%%%%%%%%%%%%%%%%
% APPENDIX
%%%%%%%%%%%%%%%%%%%%%%%%%%%%%%%%%%%%%%%%%%%%%%%%%%%%%%%%%%%%%%%%%%%%%%%%%%%%%%%
%%%%%%%%%%%%%%%%%%%%%%%%%%%%%%%%%%%%%%%%%%%%%%%%%%%%%%%%%%%%%%%%%%%%%%%%%%%%%%%
\newpage
\appendix
% Appendix
\onecolumn

\section{Target-Control Matching}
\label{app:matching}

\begin{figure*}[ht]
    % \centering
    \begin{subfigure}[t]{0.49\textwidth}
        \centering
        \includegraphics[width=\textwidth]{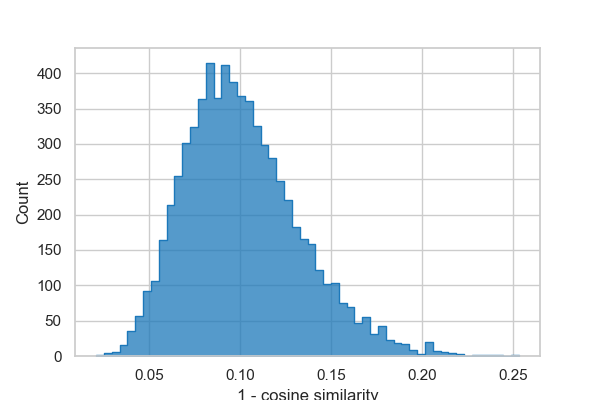}
        \caption{Cosine Distances of Matches for @\_akhaliq}
        \label{fig:scores:akhf}
    \end{subfigure}
    \hfill
    \begin{subfigure}[t]{0.49\textwidth}
        \centering        
        \includegraphics[width=\textwidth]{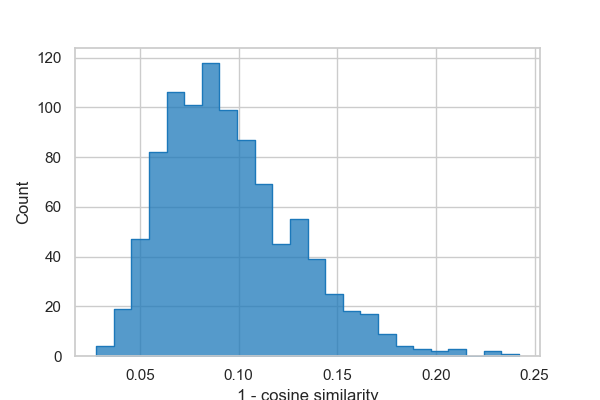}
        \caption{Cosine Distances of Matches for @arankomatsuzaki}
        \label{fig:scores:akgt}
    \end{subfigure}
    \caption{Plots of the cosine distances (1 - cosine similarity) between matched pairs for each of @\_akhaliq and @arankomatsuzaki. We note that the cosine distances are low, with the a mode of around 0.07 for both. This indicates that most pairs are very well matched in topic.}
    \label{fig:match-scores}
\end{figure*}

Qualitatively, the matched pairs are very similar in topic, almost always covering the same sub-field of research (for example, language model hallucinations). We supply a random sample of matched pairs in table \ref{tab:matched_pairs}. The distribution of scores is shown in table \ref{fig:match-scores}, and shows that the matched pairs have high cosine similarity scores. Our results are consistent across many choices of matching schemes, such as when we use more than three quantiles for binned variables, or exclude author characteristics altogether.

\begin{table*}[ht!]
    \centering
    \begin{tabularx}{\textwidth}{| l | X |}
        \hline
        
        \textbf{Score:} 0.09 & \textbf{Target Title:} Unifying Vision, Text, and Layout for Universal Document Processing \\
        & \textbf{Target Abstract:} We propose Universal Document Processing (UDOP), a foundation Document AI model which unifies text, image, and layout modalities together with varied task forma... \\
        & \textbf{Control Title:} SwinTextSpotter: Scene Text Spotting via Better Synergy between Text Detection and Text Recognition \\
        & \textbf{Control Abstract:} End-to-end scene text spotting has attracted great attention in recent years due to the success of excavating the intrinsic synergy of the scene text detection ... \\
        \hline
        
        \textbf{Score:} 0.09 & \textbf{Target Title:} reStructured Pre-training \\
        & \textbf{Target Abstract:} In this work, we try to decipher the internal connection of NLP technology development in the past decades, searching for essence, which rewards us with a (pote... \\
        & \textbf{Control Title:} Adapting BigScience Multilingual Model to Unseen Languages \\
        & \textbf{Control Abstract:} We benchmark different strategies of adding new languages (German and Korean) into the BigScience's pretrained multilingual language model with 1.3 billion para... \\
        \hline

        \textbf{Score:} 0.12 & \textbf{Target Title:} Quantifying Memorization Across Neural Language Models \\
        & \textbf{Target Abstract:} Large language models (LMs) have been shown to memorize parts of their training data, and when prompted appropriately, they will emit the memorized training dat... \\
        & \textbf{Control Title:} How Much Data Are Augmentations Worth? An Investigation into Scaling Laws, Invariance, and Implicit Regularization \\
        & \textbf{Control Abstract:} Despite the clear performance benefits of data augmentations, little is known about why they are so effective. In this paper, we disentangle several key mechani... \\
        \hline

        \textbf{Score:} 0.07 & \textbf{Target Title:} Do Language Models Know When They're Hallucinating References? \\
        & \textbf{Target Abstract:} State-of-the-art language models (LMs) are famous for"hallucinating"references. These fabricated article and book titles lead to harms, obstacles to their use, ... \\
        & \textbf{Control Title:} Retrieving Supporting Evidence for LLMs Generated Answers \\
        & \textbf{Control Abstract:} Current large language models (LLMs) can exhibit near-human levels of performance on many natural language tasks, including open-domain question answering. Unfo... \\
        \hline

        \textbf{Score:} 0.10 & \textbf{Target Title:} Singularity: Planet-Scale, Preemptive and Elastic Scheduling of AI Workloads \\
        & \textbf{Target Abstract:} Lowering costs by driving high utilization across deep learning workloads is a crucial lever for cloud providers. We present Singularity, Microsoft's globally d... \\
        & \textbf{Control Title:} Edge Impulse: An MLOps Platform for Tiny Machine Learning \\
        & \textbf{Control Abstract:} Edge Impulse is a cloud-based machine learning operations (MLOps) platform for developing embedded and edge ML (TinyML) systems that can be deployed to a wide r... \\
        \hline

        \textbf{Score:} 0.09 & \textbf{Target Title:} A Large-Scale Study on Unsupervised Spatiotemporal Representation Learning \\
        & \textbf{Target Abstract:} We present a large-scale study on unsupervised spatiotemporal representation learning from videos. With a unified perspective on four recent image-based framewo... \\
        & \textbf{Control Title:} Target-Aware Object Discovery and Association for Unsupervised Video Multi-Object Segmentation \\
        & \textbf{Control Abstract:} This paper addresses the task of unsupervised video multi-object segmentation. Current approaches follow a two-stage paradigm: 1) detect object proposals using ... \\
        \hline

    \end{tabularx}
    \caption{Randomly sampled matched pairs of papers tweeted by Komatsuzaki. Score refers to $1 - \text{cosine similarity}$, so lower scores indicate a closer match. Controls show similar topics to the target individuals across similarity levels, indicating a good match.}
    \label{tab:matched_pairs}
\end{table*}

\clearpage
\section{Additional Analysis Plots}
\label{app:plots}

\begin{figure*}[h!]
    \centering
    \begin{subfigure}[t]{0.49\textwidth}
        \centering
        \includegraphics[width=\textwidth]{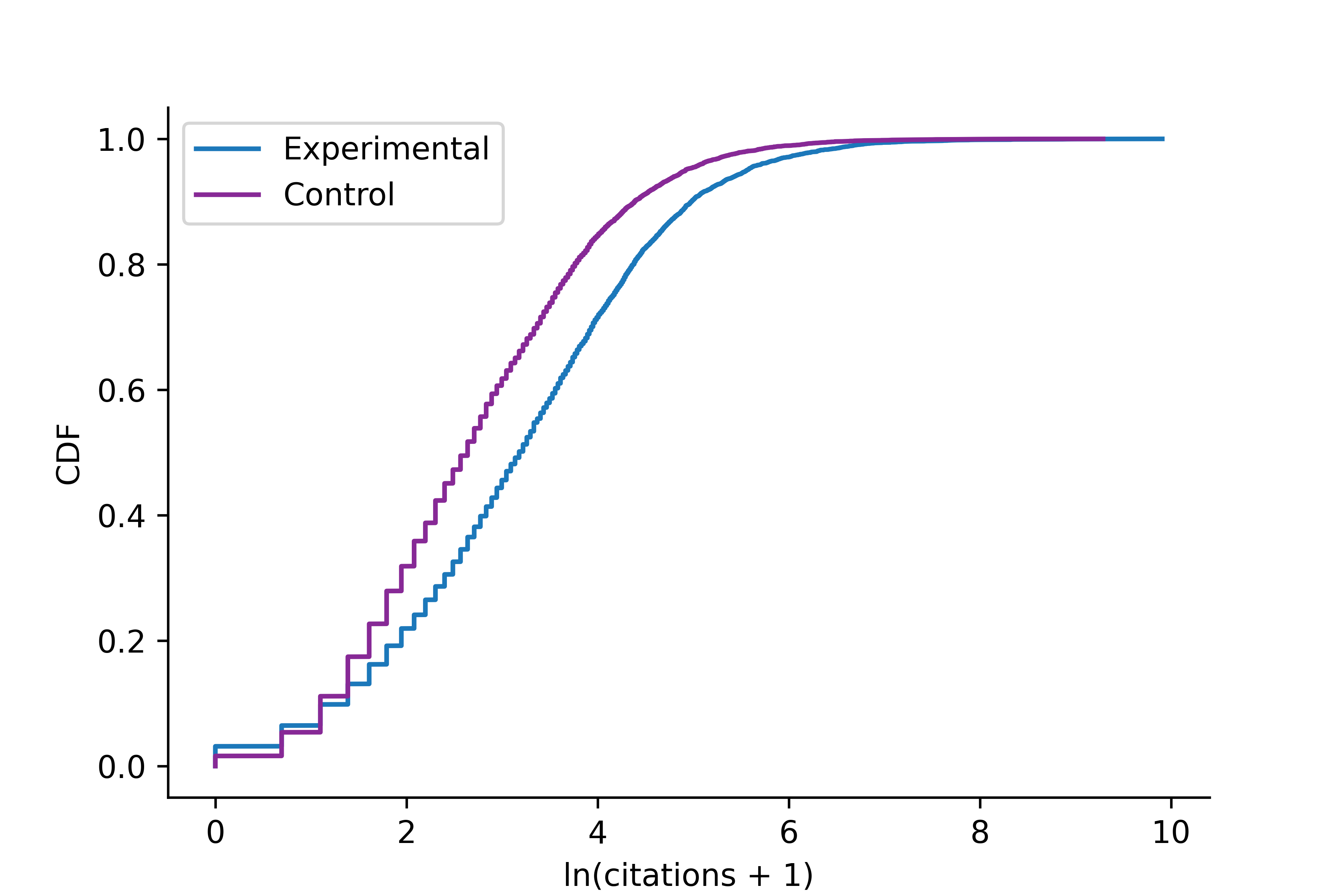}
        \caption{@\_akhaliq Dataset Cumulative Distribution Functions}
        \label{fig:cdf:akhf}
    \end{subfigure}
    \hfill
    \begin{subfigure}[t]{0.49\textwidth}
        \centering        
        \includegraphics[width=\textwidth]{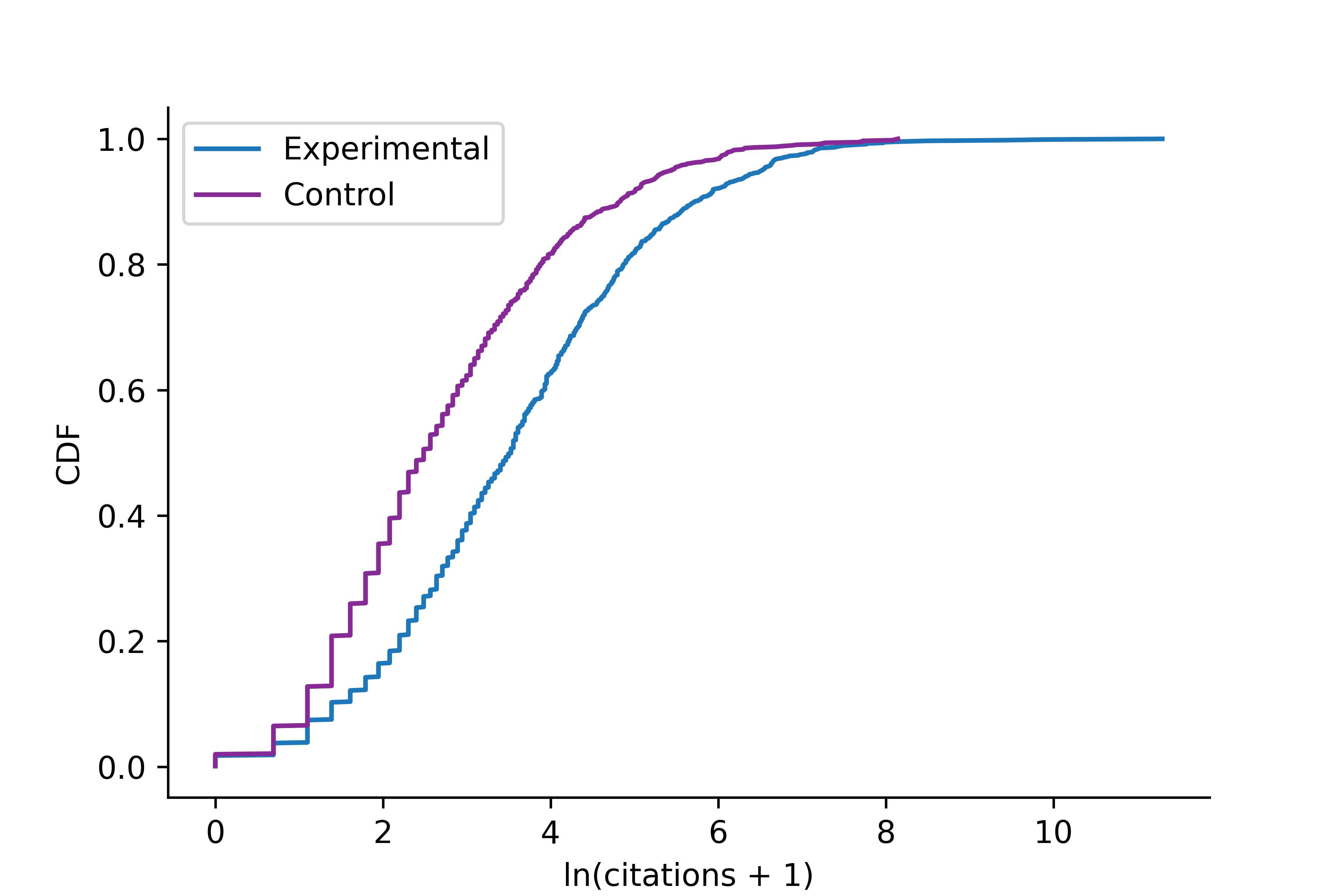}
        \caption{@arankomatsuzaki Dataset Cumulative Distribution Functions}
        \label{fig:cdf:akgt}
    \end{subfigure}
    \begin{subfigure}[t]{0.49\textwidth}
        \centering
        \includegraphics[width=\textwidth]{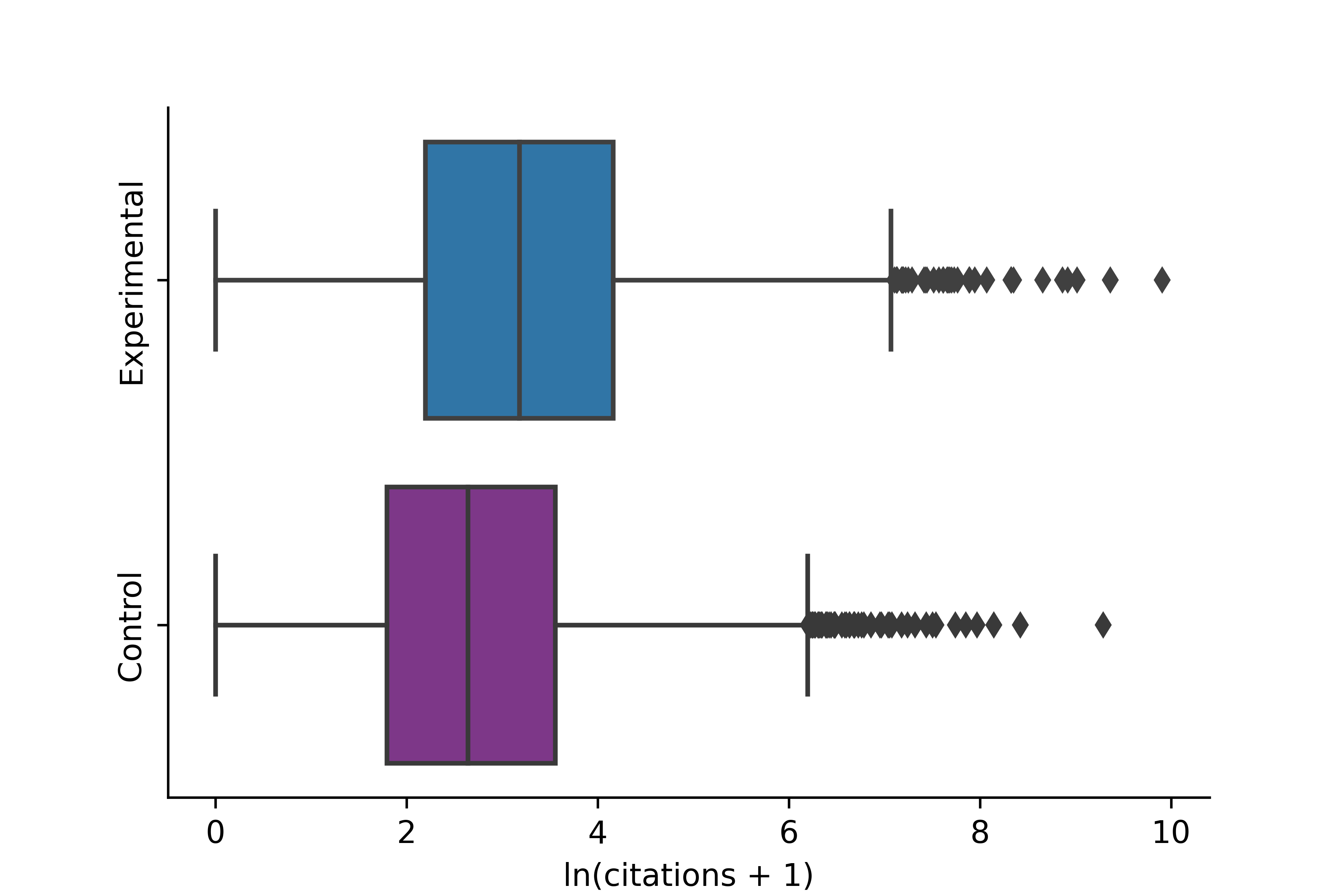}
        \caption{@\_akhaliq Dataset Box Plots}
        \label{fig:box:akhf}
    \end{subfigure}
    \hfill
    \begin{subfigure}[t]{0.49\textwidth}
        \centering
        \includegraphics[width=\textwidth]{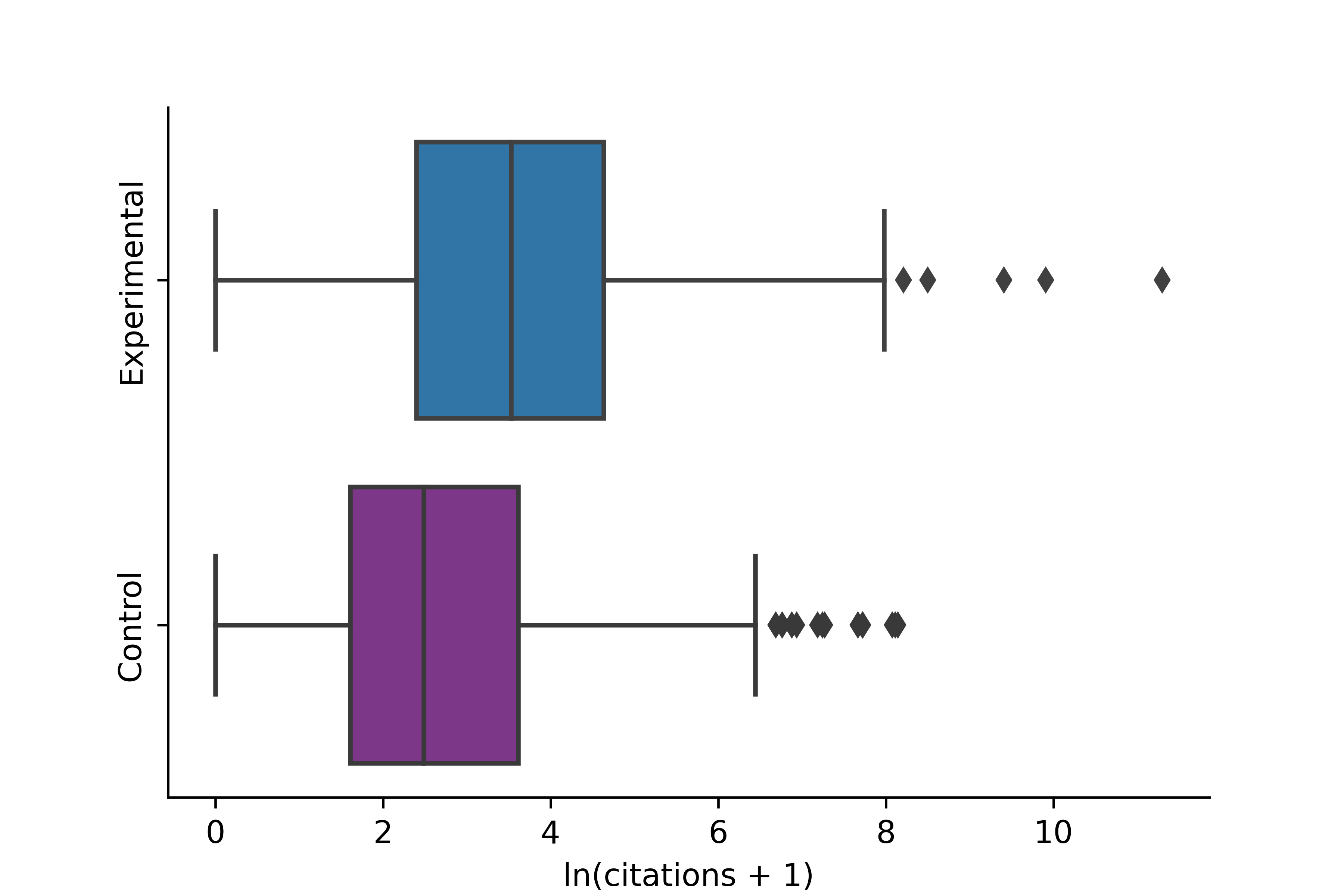}
        \caption{@arankomatsuzaki Dataset Box Plots}
        \label{fig:box:akgt}
    \end{subfigure}
    \caption{Additional plots showing the distribution of citations in the two experimental datasets and matched control samples. Citation counts are scaled with the natural logarithm using \texttt{numpy.log1p}.}
    \includegraphics[width=0.49\textwidth]{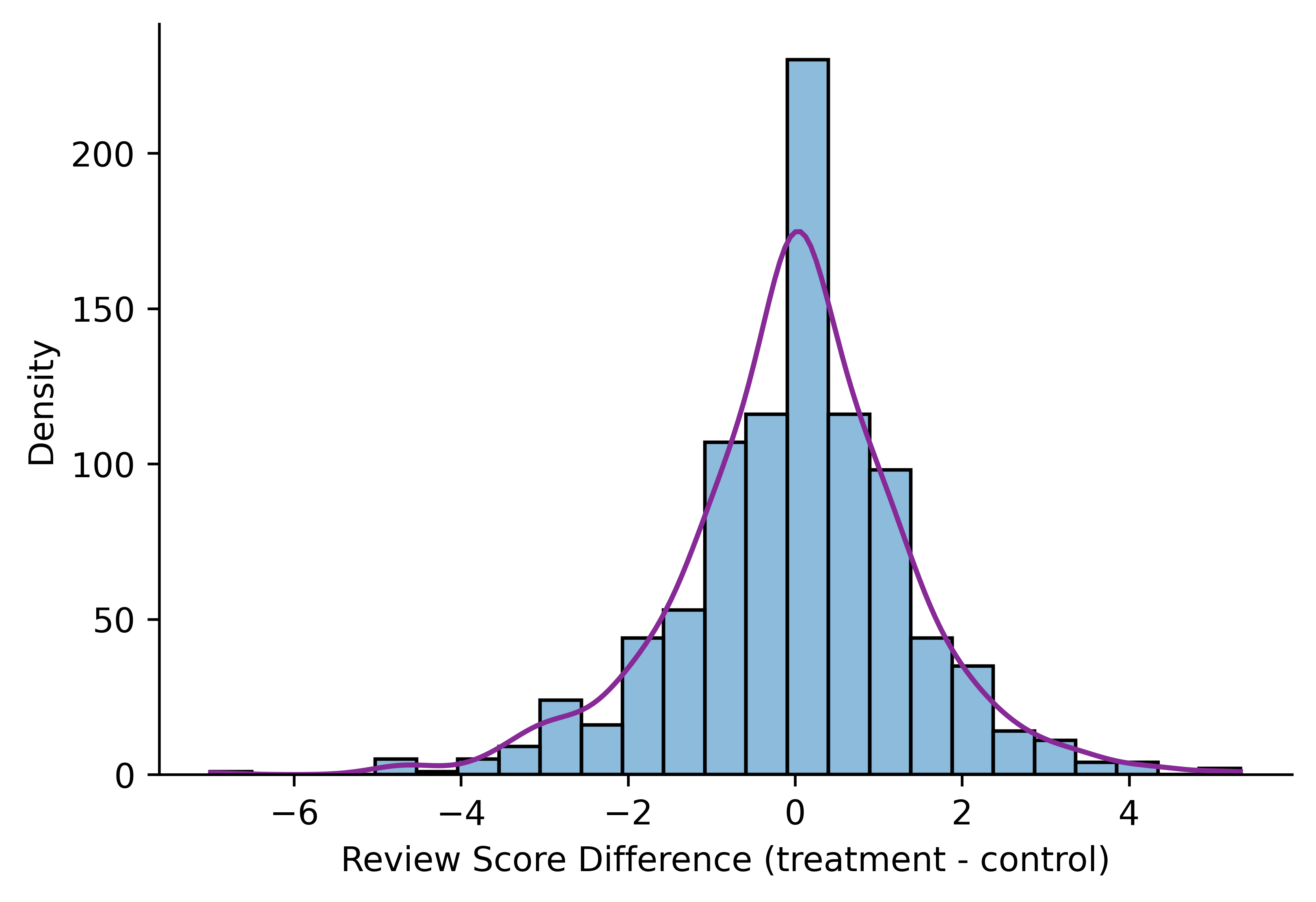}
    \caption{Histogram and kernel distribution estimate of the OpenReview score difference between experiment and control samples in the merged dataset. The mean at 0, and symmetric KDE reinforce that both review scores are from the same distribution.}
    \label{fig:app-plots}
\end{figure*}

\newpage 
\twocolumn
\section{Causal Inference}
\label{app:causalinf}

For our causal inference analysis, we draw primarily from the model, techniques, and assumptions presented in \citet{Elazar2023EstimatingTC}. In turn, they point interested readers to \citet{Imbens2015CausalIF} and \citet{Feder2021CausalII}.

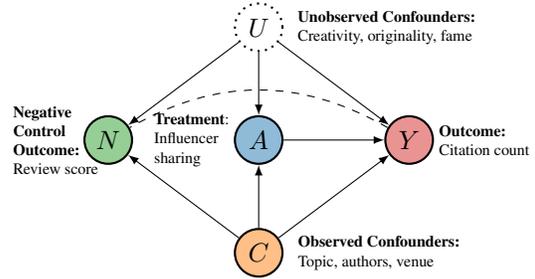
\begin{figure}[htbp]
    \centering
    \begin{center}
\begin{tikzpicture}[node distance=2.2cm,>=latex, scale=1]

\node[obs, fill=sbsblue, fill opacity=0.5, text opacity=1] (arxiv) [] at (0, 0) {$A$};
\node[obs, fill=sbsorange, fill opacity=0.5, text opacity=1] (confounder) [] at (0, -1.5) {$C$};
\node[latent] (latent) [] at (0, 1.5) {$U$};
\node[obs, fill=sbsgreen, fill opacity=0.5, text opacity=1] (nco) [] at (-2, 0) {$N$};
\node[obs, fill=sbsred, fill opacity=0.5, text opacity=1] (acceptance) [] at (2, 0) {$Y$};

\node[align=left, inner sep=0.5ex, scale=0.6] at (1.7, 1.5) {{\bf Unobserved Confounders:}\\ Creativity, originality, fame};

\node[align=left, inner sep=0.5ex, scale=0.6] at (-0.89, 0) {{\bf Treatment}: \\ Influencer \\sharing};

\node[align=left, inner sep=0.5ex, scale=0.6] at (-2.7, 0) {{\bf Negative} \\ {\bf Control} \\ {\bf Outcome:}\\ Review score};

\node[align=left, inner sep=0.5ex, scale=0.6] at (3, 0) {{\bf Outcome:} \\ Citation count};

\node[align=left, inner sep=0.5ex, scale=0.6] at (1.6, -1.5) {{\bf Observed Confounders:}\\ Topic, authors, venue};

\draw[->] (arxiv) -- (acceptance);
\draw[->] (confounder) -- (acceptance);
\draw[->] (latent) -- (acceptance);
\draw[->] (confounder) -- (arxiv);
\draw[->] (latent) -- (arxiv);
\draw[->] (latent) -- (nco);
\draw[->] (confounder) -- (nco);
\path[draw,dashed,-] (acceptance) to [out=150,in=30] (nco);

\end{tikzpicture}
\end{center}
    \caption{Causal graph of our setup. %
    $A$ and $Y$ are binary treatment and effect variables, respectively: whether a paper was shared by influencers, and whether the paper is highly cited. As we cannot measure the unobserved confounders (e.g. quality), we estimate the effect of sharing using a negative control outcome variable ($N$). Solid edges represent a directed causal effect, while dashed edges represent an association. Figure and caption modified from \citet{Elazar2023EstimatingTC}.
    }
    \label{fig:graph}
\end{figure}

\paragraph{Background. } 
In our work, we study the outcome as a binary variable $Y$, indicating if a paper is "highly-cited" ($Y=1$) or "less-cited" ($Y=0$). By introducing our treatment $A$, we can define two cases for any given paper: $Y_1$, the outcome if the paper were shared by influencers (when $A=1$), or $Y_0$, the outcome if the paper were not shared by influencers (when $A=0$). In actuality, we can only observe exactly one of these for each paper. Rather, the goal of causal inference is to estimate the change in outcome if the treatment were flipped (Figure~\ref{fig:graph}). Specifially, we hope to answer the question, \textit{does influencer sharing increase the likelihood of a paper being highly-cited?}

To accomplish this, we use the \textit{average treatment effect on treated}:
$$\text{ATET} = \mathbb{E}[Y_{1}-Y_{0}|A=1]$$
which is the expected increase in highly-cited papers when shared by influencers, for those already shared.

\paragraph{Negative Outcome Control and Difference-in-Difference. }
To debias the effects of our unobserved confounders, we will utilize the Negative Outcome Control (NOC) framework \cite{Card1993MinimumWA, Lipsitch2010NegativeCA}. Specifically, we will use Difference-in-Difference (DiD) to estimate ATET over two time periods $t\in\{0,1\}$ \cite{Sheppard1999EffectsOA}:
$$\text{ATET} = \mathbb{E}[Y_{1}(1)-Y_{0}(1)] - \mathbb{E}[Y_{1}(0)-Y_{0}(0)]$$
which can be estimated by a fitting a logistic model and bootstrapping for confidence intervals \cite{Elazar2023EstimatingTC}.

\citet{Sofer2016OnNO} establishes that NCO and DiD-based estimators are equivalent, meaning ATET can be rewritten as:
$$\text{ATET} = \mathbb{E}[Y_{1}-N_{1}] - \mathbb{E}[Y_{0}-N_{0}]$$
and solved by taking the difference across treatment groups of the difference between $Y$ and $N$.

\paragraph{Assumptions. }
To perform our causal estimate, we assume the following conditions hold:
\begin{enumerate}
    \setlength\itemsep{-0.4em}
    \item \textbf{Ignorability:} $\{Y_0, Y_1\} \perp\!\!\!\perp A | (C, U)$
    \item \textbf{Positivity:} $0 < \mathbb{P}(A = 1 | C = c, U = u) < 1$
    \item \textbf{Consistency:} $Y_a = Y^{obs} \textit{ if } A^{obs} = a$
    \item \textbf{Negative control:} $N_a = N \textit{ for } a = 0, 1$
    \item \textbf{Additive Equi-confounding: } \\$\mathbb{E}[Y_a(1)-Y_a(0)|U,A=a,C]  \\=\mathbb{E}[Y_a(1)-Y_a(0)|A=a,C] \textit{ for } A=0,1$
\end{enumerate}

In other words, \textit{ignorability} assumes $(C, U)$ are the only confounders with an effect on the outcome; \textit{positivity} means each paper has a non-zero chance of being and not being influencer-shared; \textit{consistency} establishes that the potential outcomes $Y_a$ and observed outcomes $Y^{obs}$ agree for a given treatment level $A^{obs}$; \textit{negative control} assumes that the NCO is not causally affected by the treatment $A$; \textit{additive equi-confounding} relies on a linear (or logistic) outcome model for $U$.

\paragraph{Review Scores as a Negative Control Outcome. }
The goal of the Negative Outcome Control (NOC) framework \cite{Card1993MinimumWA, Lipsitch2010NegativeCA} is to find a Negative Control Outcome (NCO) variable that is affected by the same confounders $(C, U)$ as the outcome $Y$, but not the treatment. This way, we can de-bias the effects of confounders not captured by the metadata and text embeddings collected from S2. For this, we choose the average conference review score for each paper, collected as described in Section~\ref{sec:data}. We believe that this sub-sample is representative for the following reasons: (1) influencers often share preprint works before conference acceptance or proceedings are released; (2) our matching accounts for publication venue as a confounder; (3) using bootstrapping on the three tests from Table~\ref{tab:2-sample_tests}, we find there is not significant evidence to reject that the sub-sample and full dataset citations come from different distributions ($p$-value $>0.05$).

\paragraph{Matching and Causal Inference. }
\citet{Elazar2023EstimatingTC} use tripartite matching on a number of covariates \cite{Zhang2021MatchingOS}, and only sampled from one venue. To control the topic covariate, they use SPECTER embeddings to build 20 topic clusters and minimize distance for numerical variables (author citations, institution rank, etc.)

Instead, we group the treated and control samples based on a mix of exact matching (venue, year, open access) and quantile binning (number of authors, h-index, and author cites). From these groups, we match samples by minimizing the $L_2$ difference of text embeddings. This way, we can most accurately capture increased nuance between "viral" topics, techniques, and models. 

Additionally, we use first-author information, rather than both minimum and average, to increase our group sizes and improve the quality of topic matches while maintaining descriptive author details. 

\section{Limitations}
\label{app:limitations}

While our study provides valuable insights into the role of influencers in the machine learning (ML) community, there are several limitations that must be acknowledged:

\textbf{Binary Gender Analysis:} Our gender prediction utilizes the AMiner Gender Prediction API, which combines the Facebook Generated Names List \cite{Tang2011NamesGenderFacebook} with precision and recall of 95\% and 81\%, and MagicFG \cite{Gu2016WebUP} with precision and recall of 93\% and 94\% respectively. Exact numbers for the combined API are not published by the creators but can reasonably be assumed to be higher than each individual method. 

Additionally, the gender prediction API only outputs binary genders (male and female), limiting our ability to capture the full spectrum of gender identities. Consequently, our findings may not accurately represent the diversity of gender identities within the ML community. 

\textbf{Accuracy with Non-Western Names:} The gender prediction API may exhibit lower accuracy when analyzing non-Western names \cite{Gu2016WebUP}. As a result, the study's conclusions regarding gender disparity might not fully capture the global diversity of our samples.

\textbf{Self-Reported Affiliations:} The affiliations in our dataset are self-reported, which can lead to inconsistencies such as misspellings or missing location information. This limitation affects the accuracy of our geographical analysis, as the data may not accurately reflect the true affiliations or locations of the authors. Consequently, our observations regarding geographical concentration and diversity must be interpreted with caution. Additionally, this limitation posed as a barrier to investigating institutions as an observed confounder for causal inference. We invite future studies with more complete affiliation data to investigate this.

\textbf{Lack of Randomized Control Trial:} The study did not include a randomized control trial, often considered the gold standard for establishing causality. Without this, we cannot conclusively determine whether the patterns observed are directly attributable to the influencers' activities or are coincidental. Though we attempt to remedy this through causal inference, such techniques are contingent upon the validity of our assumptions.

%%%%%%%%%%%%%%%%%%%%%%%%%%%%%%%%%%%%%%%%%%%%%%%%%%%%%%%%%%%%%%%%%%%%%%%%%%%%%%%
%%%%%%%%%%%%%%%%%%%%%%%%%%%%%%%%%%%%%%%%%%%%%%%%%%%%%%%%%%%%%%%%%%%%%%%%%%%%%%%

\end{document}